\documentclass{aa}  
\usepackage{graphicx}
\usepackage{graphbox} % it depends on "graphicx" package
\usepackage{eqnarray,amsmath}
\usepackage{mathtools}
\usepackage{multicol}
\usepackage[hypcap]{caption}
\usepackage{txfonts}
\usepackage{subcaption}
\usepackage{arydshln}
\usepackage{hyperref}
\usepackage{xcolor}
\usepackage{ulem}
\usepackage{cancel}
\graphicspath{{./figures/}} % Location of the graphics files

\begin{document} 
\titlerunning{Evolution of stars with $60$ and $200$ $M_\odot$: predictions for WNh stars in the Milky Way.}
\title{Evolution of stars with $60$ and $200$ $M_\odot$:\\
predictions for WNh stars in the Milky Way.}
\authorrunning{Gormaz-Matamala et al.}
\author{A. C. Gormaz-Matamala\inst{1,2}%,2,3}
\and
A. Romagnolo\inst{1,3}
\and
K. Belczynski\inst{1}\thanks{Deceased on 13th January 2024.}
}
\institute{
Nicolaus Copernicus Astronomical Center, Polish Academy of Sciences, Bartycka 18, 00-716 Warsaw, Poland
\and
Astronomický ústav, Akademie věd Ceské republiky, Fričova 298, 251 65 Ondřejov, Czech Republic
\and
Department of Astronomy and Astrophysics, University of California San Diego, La Jolla, California 92093, USA
\\
\email{alex.gormaz@asu.cas.cz}
}
\date{}

\abstract % 5 {} token are mandatory
{Massive stars are characterised by powerful stellar winds driven by radiation, and therefore mass-loss rate plays a crucial role in their evolution.}
{We study in detail the evolution of two massive stars (a classical massive star and a very massive star) at solar metallicity ($Z=0.014$), by calculating their final masses, radial expansion, and chemical enrichment, at their H-core, He-core and C-core burning stages prior to the final collapse.}
{We run evolutionary models for initial masses $60$ and $200$ $M_\odot$, using \texttt{MESA} and the Geneva-evolution-code (\textsc{Genec}).
For the mass loss, we adopt the self-consistent m-CAK prescription for the optically thin winds of OB-type stars, a semi-empirical formula for H-rich optically thick wind of WNh stars, and a hydrodynamically consistent formula for the H-poor thick wind of classical Wolf-Rayet stars.
The transition from thin to thick winds is set at $\Gamma_\text{e}=0.5$.}
{The unification of the initial setup for stellar structure and wind prescription leads into very similar $M_\text{bh}$ for both \textsc{Genec} and \texttt{MESA} codes, but both codes predict different tracks across the HRD.
For the $60\,M_\odot$ case, the \textsc{Genec} model predicts a more efficient rotational mixing and more chemically homogeneous evolution, whereas the \texttt{MESA} model predicts a large radial expansion reaching the LBV phase.
For the $200\,M_\odot$ case, differences between both evolution codes are less relevant because their evolution is dominated by wind mass loss with a weaker dependence on internal mixing.}
{The switch of the mass-loss prescription based on the Eddington factor instead of the removal of outer layers, implies the existence of WNh stars with a large mass fraction of hydrogen at the surface ($X_\text{surf}\ge0.3$) formed from initial masses of $\gtrsim60\,M_\odot$.
These stars are constrained in a $T_\text{eff}$ range of the HRD which corresponds to the main sequence band, in agreement with the observations of Galactic WNh stars at $Z=0.014$.
While our models employ a fixed $\Gamma_\text{e,trans}$ threshold for the switch to thick winds, rather than a continuous thin-to-thick wind model, the good reproduction of observations during the main sequence supports the robustness of the wind model upgrades, allowing its application to studies of late-stage stellar evolution before core collapse.}

\keywords{Stars: early-type -- Stars: evolution -- Stars: mass-loss -- Stars: winds, outflows -- Stars: Wolf-Rayet}
\maketitle

%_____INTRODUCTION_______________________________________________________________________________
%\begin{multicols}{1}
\section{Introduction}
	The detection of the first source of gravitational waves \citep[GW,][]{abbott16a,abbott21} opened the door for a large number of studies dedicated to find more GW sources in the Universe, namely \textit{double compact objects} (hereafter DCO) mergers.	
	Rates of these DCO merger events depend on considered conditions for the binary system concerning the orbital separation, the formation (or not) of a common envelope (CE), as well as the initial assumptions for RLOF and their timescales  \citep{ivanova13b,olejak21}.
	Moreover, these modes are intimately linked with the single evolution of the stars in a binary system, which is crucial to determine the final remnant masses \citep{yungelson08,belczynski10a,yusof13,spera17,bavera23,martinet23,romagnolo24} and the maximum radial expansion that the stellar components can reach during their evolution \citep{agrawal20,romagnolo23}, reason why accurate evolution models are necessary to constrain the theoretical rate of the DCO merging events.

	Stellar black holes are the final fate of massive stars with $M_\text{zams}\gtrsim25\,M_\odot$ \citep{heger03}.
	Even though the conditions of the death as supernovae also constrain the mass of the final remnant \citep{fryer22,olejak22}, the stellar mass prior to the supernova explosion is explicitly dependent of the mass-loss rate (denoted as $\text{d}M/\text{d}t$ or simply $\dot M$) carried out by the strong winds during the evolution of a massive stars \citep{vink21b}.
	Because stellar winds of hot massive stars are driven by line-force \citep{lucy70,cak,abbott82}, these $\dot M$ are directly dependent on metallicity.
	This explains the predominance of more massive black holes in less metallic environments \citep{farrell21,vink21a}, although exceptions due to unusually weaker winds can also be theoretically described \citep{belczynski20a}.
	Moreover, the strength of the stellar winds varies along the evolution of the star, which is dominated during the main sequence (H-core burning stage) by line-driven optically thin winds for OB-type stars \citep{puls08}, then eventual strong outflows might stem during the LBV stage \citep{humphreys16}, and finally we find optically thick winds of Wolf-Rayer (WR) stars for the more evolved stages \citep{crowther07,grafener08}.
	Optically thick winds are naturally denser, and therefore the mass loss during the WR phase is higher in comparison with OB-type phase.

	Recent theoretical \citep{kk17,kk18,alex19,alex22a,bjorklund21} and empirical studies \citep{hawcroft21,hawcroft23} have found theoretical values for the $\dot M$ of massive stars to be in the order of $\sim3$ times lower than the standard mass-loss recipe from \citet[][hereafter V01]{vink01}.
	Such difference in the value of mass loss is enough to alter the evolution of stars with $M_\text{zams}\gtrsim20\,M_\odot$ \citep{meynet94}.
	Adopting new winds, massive stars will retain more mass during their main sequence phase, thus being \text{bigger} and more luminous \citep{alex22b,bjorklund23}.
	Besides, massive stars keep larger rotational velocities for a more extended time, because they lose less angular momentum \citep{alex23a}.
	Even though these changes in the stellar evolution have been applied mostly for the early H-burning stages, they also impact the further evolution of the subsequent stages \citep{alex24a,josiek24}.
	Moreover, considering that also the mass-loss prescription for WR winds has been revisited \citep{bestenlehner20,sander20} as well as its impact on the stellar evolution \citep{higgins21}, the entire panorama of the stellar evolution is being upgraded.
	
	In \citet[][hereafter Paper I]{romagnolo24}, we calculate the remnant mass for a set of initial masses from $10$ to $300$ $M_\odot$ for $Z=0.014$, based on stellar evolution models adopting the state-of-the-art physical ingredients such as Tayler-Spruit dynamo, larger core-convective overshooting, and new mass-loss recipes.
	As a result, we found a maximum black hole mass of $M_\text{bh,max}=28.3\,M_\odot$ formed from stars with $M_\text{zams}\gtrsim250\,M_\odot$, larger than the previous $M_\text{bh,max}\simeq20\,M_\odot$ from \citet{belczynski10a} but quite below the maximum BH masses predicted by \citet[][$\simeq35\,M_\odot$ at $M_\text{zams}=75\,M_\odot$]{bavera23} and \citet[][$\simeq42\,M_\odot$ at $M_\text{zams}=180\,M_\odot$]{martinet23}.
	Even though the recently discovered largest BH in the Milky Way has a mass of $M_\text{bh,max}=32.7\pm0.70\,M_\odot$ \citep{gaia24}, the low metallicity of its star companion suggest that this black hole was formed from a metal poor massive star and thus our predicted $M_\text{bh,max}$ remains valid.

	Moreover, the new wind prescription introduced in Paper I entail more implications concerning the evolution of massive stars.
	Standard evolution models \citep{ekstrom12,choi16} adopted the switch from OB-type to WR-type winds once the stellar surface removes the majority of its hydrogen.
	Nevertheless, theoretical studies have found that the transition from optically thin to thick winds during the evolution of massive stars is more due to the proximity to the Eddington limit rather than due to the removal of the outer envelopes \citep{vink11,grafener11}, thus altering the expected lifetimes and total amount of mass removed at each evolutionary stage.
	Because of this, in Paper I we set the switch between optically thin to optically thick winds according to the proximity to the Eddington limit instead of based on the chemical composition of the stellar outer envelope.
	As a consequence, the most massive stars develop thick winds already during their earliest evolution stages, altering their tracks over the HR diagram and predicting the existence of WR stars with hydrogen abundance $X_\text{surf}\ge0.3$ in their surfaces, the so-called WNh stars \citep{martins08,crowther10}.

	In this work, we study in detail the impact of the already mentioned upgrades on the evolution of two stars from Paper I, with special emphasis on the existence of H-rich WR stars.
	For that purpose, we run evolution models for the stars with initial masses of $60$ and $200$ $M_\odot$ at solar metallicity ($Z=0.014$), following \citet{asplund09}. 
	We introduce our mass-loss prescription in Sec.~\ref{masslossupgrade}, together with the description of the switch from optically thin to thick winds at $\Gamma_\text{e}=0.5$.
	The resulting evolution tracks over the HR diagram for our models are detailed in Sec.~\ref{mainresults}, whereas the observational diagnostics comparing with spectroscopic analyses are shown in Sec.~\ref{diagnostics}.
	Finally, conclusions are summarised in Sec.~\ref{conclusions}.

%_____UPGRADE OF MASS LOSS RECIPES
\section{Ingredients for stellar evolution models}\label{masslossupgrade}
%_____Upgrades on the inner stellar structure
\subsection{Stellar evolution codes}\label{evolutioncodes}
	Same as in Paper I, we perform our stellar evolution models using the codes {\tt MESA} \citep[Modules for Experiments in Stellar Astrophysics, ][]{paxton11,paxton13,paxton15,paxton18,paxton19,jermyn23} and the Geneva-evolution-code \citep[][hereafter \textsc{Genec}]{eggenberger08}.
	For step-overshooting, we adopt a value for $\alpha_\text{ov}$ of $0.5$, in agreement with the most recent studies on convective cores \citep{martinet21,scott21,baraffe23}.
	Concerning angular momentum transport, we include for both {\tt MESA} and \textsc{Genec} codes the Tayler-Spruit dynamo \citep{tayler73,spruit02,heger05}, whereas we replace the treatment for the convective layers from Schwarzchild to Ledoux's criterion \citep{ledoux47}.
	For the convective outer layers, \texttt{MESA} adopts a mixing-length parameter of $\alpha_\text{MLT}=1.82$, whereas \textsc{Genec} keeps $\alpha_\text{MLT}=1.6$.
	In order to reduce superadibaticity in regions near the Eddington limit we use the \texttt{use\_superad\_reduction} method, which is stated to be a more constrained and calibrated prescription than MLT++ in \texttt{MESA} \citep{jermyn23}.
	In particular, this difference in the superadiabacity treatment will be relevant during the VMS evolution across the He-core burning stage \citep{romagnolo24b}, as we observe in Sec.~\ref{heliumcoreevolution}.
	
\subsection{Mass loss for massive and very massive stars}\label{masslossupgrade}
	Evolutionary tracks of massive stars adopting lower $\dot M$ are in rule with the observed mass loss of clumped winds \citep{bouret05,muijres11,surlan13}, as well as with more detailed calculations for the wind structure of O-type stars by self-consistently solving radiative transfer equations at full NLTE conditions \citep{sander17,sundqvist19,alex21}.        	
	However, this scenario can hardly be extended to higher masses such as VMS, whose mass-loss rates values even larger than the given ones by V01 \citep{vink11}.
	Their stellar winds are so strong that they develop WR-type spectral profile with broad emission lines before the end of the H-core burning stage (WNh stars).
	The limit between VMS and ``regular'' massive stars is normally set around $M_\text{zams}\sim100$ $M_\odot$, although this value may vary by means of rotation, metallicity, and specially mass-loss input \citep{yusof13,kohler15}.

	Evolution of VMS, revisiting the transition between formal O-type to WNh stars, has been just recently analysed by \citet{grafener21} and \citet{sabhahit23}.
	Despite this, standard evolution codes still assume the transition of the wind regime based on the surface composition (when the hydrogen mass fraction becomes $X_\text{surf}\le0.3$ for \textsc{Genec} and  $X_\text{surf}\le0.4$ for \texttt{MESA}) the mass-loss recipe switches from the prescription for optically thin V01 to optically thick winds according to \citet[][hereafter GH08\footnote{Nevertheless, evolution models used to both V01 and GH08 to describe mass loss from WR winds, depending on which recipe provided the largest $\dot M$ at certain point \citep{yusof13}.}]{grafener08} in \textsc{Genec}, whereas \texttt{MESA} uses the prescription of \citet{nugis02} for WR winds.
	However, spectroscopic diagnostics have found WNh spectral features for VMS with surface hydrogen abundances as high as $X_\text{surf}\simeq0.7$ in the Milky Way \citep{martins23} and in the Large Magellanic Cloud \citep{tehrani19,martins22}, thus strengthening the idea that the transition to thick winds should be based on another criteria apart from surface chemical composition.

%_____Proximity to the Eddington limits
\subsubsection{Winds in the proximity to the Eddington limit}\label{eddingtonlimit}
	Mass loss of VMS is directly dependent on the Eddington factor, and the relationship quickly escalates when the Eddington limit is approached \citep{vink11,vink12}.
	The Eddington limit defines the limit for the hydrostatic stability of a star, and it corresponds when the outwards radiative force equals to the inwards gravitational force\footnote{Notice that, because of this definition, the Eddington limit only applies for the stellar surface} \citep{langer97}.
	This limit is better illustrated by the \textit{classical} Eddington factor, which considers only the electron scattering for the radiative processes
	\begin{equation}\label{Gammae}
		\Gamma_\text{e}=\frac{L_*\kappa_\text{e}}{4\pi cGM_*}\;,
	\end{equation}
	with $\kappa_\text{e}$ being the electron scattering opacity.
	Assuming a fully ionised plasma, the Eddington factor is usually expressed as a function of stellar mass, luminosity and hydrogen abundance
	\begin{equation}\label{logGammaEdd}
		\log\Gamma_\text{e}=-4.813+\log(1+X_\text{surf})+\log(L_*/L_\odot)-\log(M_*/M_\odot)\;,
	\end{equation}
	i.e., the classical Eddington factor depends only on the fundamental stellar parameters.
	
	Numerical models found a kink in the slope of the $\dot M\propto\Gamma_\text{e}$ dependence at $\Gamma_\text{e}\simeq0.7$, after which the mass-loss scales with the Eddington factor in agreement with the relationship found for \citet{grafener08} for \text{WNL} stars.
	Indeed, $\Gamma_\text{e}=0.7$ was already used as a transition criterion to switch from O-type to WR stars in evolution models \citep{chen15}.
	Posteriori observational diagnostics on very massive stars found that wind clumping and uncertainties affects the accuracy to determine the kink in the $\Gamma_\text{e}$ space, which can be set as low as $\Gamma_\text{e}\simeq0.3$ \citep{bestenlehner14}, even though this value for the Eddington factor strongly depends on the mass estimations derived from the $L/M$ relations from \citet{grafener11}.
	A more theoretically detailed study is provided by \citet{sabhahit23}, where the switch on the wind regime is directly correlated with the wind efficiency $\eta=\dot M\varv_\infty/(L/c)$ at different metallicities.
	In this case, the mass loss of the initial O-type phase is described by the V01 instead of more recent recipes \citep[e.g.][]{kk17,bjorklund21,alex22b}.

	Instead, we aim for a transition point between thin and thick winds, independent not only of the initial mass-loss rate adopted but also independent on the wind momentum.
	For this, we based our transition in the upper limit of validity of our m-CAK prescription \citep{alex22a}, where both $\dot M$ and the wind velocity profile $\varv(r)$ are self-consistently calculated together with the line-force parameters \citep{alex19}.
	\citet{bestenlehner20} extended the theoretical calculation of $\dot M$ according to the CAK theory \citep{cak} for regimes closer to the Eddington limit, finding also a transition $\Gamma_\text{e,trans}$ beyond the which the slope of the relation $\dot M\propto\Gamma_\text{e}$ become steeper, in line with the observations of WNh stars exhibiting thick winds from \citet{bestenlehner14}.
	This $\Gamma_\text{e,trans}$ ranges from $\sim0.46$ to $\sim0.48$, depending on the value of the CAK line-force parameter\footnote{Do not confuse this parameter $\alpha$, which is one of the line-force parameters from CAK theory \citep{cak,abbott82}, with the $\alpha_\text{ov}$ for the convective core overshooting.} $\alpha$.
	For $\alpha\sim0.4-0.6$, which are the values found from the self-consistent line-force parameters found by \citet{alex22b} for stars with masses in the range of $25$ to $120$ $M_\odot$ and metallicities from $Z=0.014$ (solar) to $=0.002$ (SMC), the value of the transition Eddington factor keeps relatively constant around $\Gamma_\text{e,trans}\simeq0.48$.
	Therefore, given that our new prescription for optically thin winds is based on the m-CAK theory, we establish $\Gamma_\text{e}=0.5$ as the transition point between thin and thick winds for our evolution models.
	The thickness in this case means the dominance of the electron scattering on the radiative acceleration, above the line-driven.
	We will discuss the physical implications of this transition in Section~\ref{selfconsistent_gamma05}.

%_____Thin and thick winds for the H-core burning stage
\subsection{Mass-loss recipes for thin and thick winds}
\subsubsection{Mass loss for the H-core burning stage}\label{newmdots}
	For thin winds ($\Gamma_\text{e}\le0.5$), we use the mass-loss prescription from \citet{alex22b,alex23a}
	\begin{align}\label{mdotalex22bnew}
		\log\dot M_\text{thin}=&-40.314 + 15.438\,w + 45.838\,x - 8.284\,w\,x \nonumber\\
		&+ 1.0564\,y -  w\,y / 2.36 - 1.1967\,x\,y\nonumber\\
		&+z\times\left[0.4+\frac{15.75}{M_*/M_\odot}\right]\;,
	\end{align}
	where $\dot M_\text{thin}$ is in $M_\odot$ yr$^{-1}$; and where $w$, $x$, $y,$ and $z$ are defined as
	\begin{equation}
		w=\log \left(\frac{T_\text{eff}}{\text{kK}}\right)\;,\nonumber\\
		x=\frac{1}{\log g}\;,\\
		y=\frac{R_*}{R_\odot}\;,\\
		z=\log \left(\frac{Z_*}{Z_\odot}\right)\;.
	\end{equation}

	As mentioned in \citet{alex23a}, the dependence between mass loss and metallicity is not constant but it intrinsically depends on the mass (and the luminosity), because the radiative acceleration for more massive and luminous stars is mostly dominated by the continuum over the line-acceleration \citep[see also][]{kk18}.

	The range of validity of Eq.~\ref{mdotalex22bnew} is for $T_\text{eff}\ge30$ kK and $\log g\ge3.2$.
	Beyond these values, we return to V01 by default, even though the $\dot M$ for stars in the range of $10-30$ kK needs to be eventually also decreased \citep{deburgos24}.
	For the cases when $T_\text{eff}\le10$ kK, mass loss is described by \citet{dejager88}.

	For thick winds ($\Gamma_\text{e}>0.5$), since we establish the transition from the theoretical analysis of \citet{bestenlehner20}, we use their $\dot M(\Gamma_\text{e})$ recipe with the calibration performed by \citet{brands22}
	\begin{equation}\label{mdotbrands22}
		\log\dot M_\text{B22}=-5.19+2.69\log(\Gamma_\text{e})-3.19\log(1-\Gamma_\text{e})\;.
	\end{equation}

	It is easy to see that, for $\Gamma_\text{e}=0.5$ Eq.~\ref{mdotbrands22} becomes $\log\dot M\simeq-5.04$.
	This value is very close of the `transition' mass-loss rates empirically determined for the Arches cluster by \citet{vink12} and for 30 Dor by \citet{sabhahit23}, i.e., the value for $\dot M$ over the which the stars of the mentioned clusters exhibit a WNh spectrum.
	This implies that both approaches \citep[][and this current study]{sabhahit23} find the transition mass-loss rate between thin and thick winds at roughly the same value.
	Moreover, because this formula was calibrated for the VMS in the R136 star cluster at the LMC, with metallicity $Z=0.006$ \citep{eggenberger21}, we add an extra term with the same metallicity dependence as for thin winds.
	Thus
	\begin{equation}\label{mdotthick}
		\log\dot M_\text{thick}=\log\dot M_\text{B22}+\log \left(\frac{Z_*}{Z_\text{LMC}}\right)\times\left[0.4+\frac{15.75}{M_*/M_\odot}\right]\;,
	\end{equation}

	Hence, despite we are using the same $\Gamma_\text{e,trans}$ for all metallicities, mass loss for optically thick winds incorporates a metallicity dependence which is consistent with the intrinsic dependence on mass/luminosity.
	Even though calibration from \citet{brands22} matches with the formula of \citet{kk18} (and subsequently with their intrinsic metallicity dependence on luminosity), we adopt the intrinsic metallicity dependence on mass introduced by \citet{alex23a} to keep coherence between our adopted thin and thick winds.

	Analogous to Eq.~\ref{mdotalex22bnew} for thin winds, Eq.~\ref{mdotthick} is valid only for $T_\text{eff}\ge30$ kK.
	For thick winds at cooler temperatures we return to V01, % since that recipe was shown to fit the aforementioned transition mass loss \citep{vink12}.
	thus their predicted bi-stability jump is still being implemented for all the temperature range between $10$ and $30$ kK, for both $\dot M_\text{thin}$ and $\dot M_\text{thick}$.

%_____Mass loss during He-core burning stage
\subsubsection{Mass loss for advanced evolutionary stages}
	The Eq.~\ref{mdotthick} from \citet{bestenlehner20} and \citet{brands22} is valid for H-rich WR stars, either for the H-core or the He-core burning stages.
	Once the star has depleted the most part of its hydrogen ($X_\text{surf}\le10^{-7}$), the mass-loss rate follows the formula derived from the hydrodynamically consistent models for WR stars from \citet{sander20a}, whose final formula $\dot M\sim\Gamma_\text{e}$ is given by \citet{sander20}
	\begin{equation}\label{wrsv20}
		\log\dot M=a\log[-\log(1-\Gamma_\text{e})]-\log 2\left(\frac{\Gamma_\text{e,d}}{\Gamma_\text{e}}\right)^{c_{d,b}}+\log\dot M_\text{off}\;,
	\end{equation}
	where
	$$a=2.932\;,$$
	$$\Gamma_\text{e,b}=-0.324*\log \left(\frac{Z_*}{Z_\odot}\right)+0.244\;,$$
	$$c_{d,b}=-0.44*\log \left(\frac{Z_*}{Z_\odot}\right)+9.15\;,$$
	$$\log\dot M_\text{off}=0.23*\log \left(\frac{Z_*}{Z_\odot}\right)-2.61\;.$$

	We do not incorporate the latest improvements to this formula from \citet{sander23}, because they add extra terms dependent on a quantity $T_*$ which is not equivalent to the effective temperature $T_\text{eff}$, and thus such implementation is currently beyond the capacities of both codes \texttt{MESA} and \textsc{Genec}.
	The only condition of validity for Eq.~\ref{wrsv20} is the criterion $X_\text{surf}\le10^{-7}$ (H-poor WR star), regardless of the Eddington factor of luminosity.
	Finally, we also mention that all the wind recipes considered here are thought for single stellar evolution, and for that reason we do not include wind recipes for peculiar scenarios coming from binary interaction, such as stripped low-mass helium stars \citep{vink17} or O-type stars with He-core burning \citep{pauli22}.

%_____Treatment of rotation
\subsection{Treatment of rotation}\label{rotationmethod}
	The most important difference concerning rotation between \texttt{MESA} and \textsc{Genec} is the standardisation.
	\texttt{MESA} sets stellar rotation as a fraction of the critical \textit{angular} rotational velocity, $\omega$, needed to disrupt the star due to centrifugal forces \citep{paxton13}, i.e.\footnote{In this work, we denote $\Omega$ to the ratio between rotational velocity and critical velocity following the formalism of the line-driven theory \citep{michel04b,puls08,venero24}, whereas for the angular velocity we use the letter $\omega$.}
	\begin{equation}\label{omega_crit}
		\Omega_\text{MESA}=\frac{\omega}{\omega_\text{crit}}\;,\;\;\text{with }\omega_\text{crit}^2=\frac{GM}{R^3}\left(1-\Gamma_\text{Edd}\right)\;,
	\end{equation}
	where $\Gamma_\text{Edd}$ is the total Eddington factor where $\kappa$ is actually the Rosseland mean opacity, calculated as a mass-weighted average in a user-specified optical depth range.

	On the other hand, \textsc{Genec} sets stellar rotation as a fraction of the \textit{linear} rotational velocity ($\Omega_\textsc{Genec}=\varv/\varv_\text{crit}$).
	However, these two fractions are not equivalent because a rotating star loses its spherical shape and becomes oblate, thus meaning that the polar radius and the equatorial radius are not longer the same.
	Even though both codes incorporate oblate effects, the formal difference between linear and angular velocity is not considered in \texttt{MESA} \citep{choi16} but only in \textsc{Genec}, by means of the Roche-model \citep{ekstrom08,georgy11}.
	According to this, the ratios for the critical angular and linear velocities are related in \textsc{Genec} as
	\begin{equation}\label{vvcrit}
		\frac{\varv}{\varv_\text{crit}}=\left[\frac{\omega}{\omega_\text{crit}}2\left(\frac{R_\text{eq}}{R_\text{pol}}-1\right)\right]^{1/3}\;,\;\;\;\text{with }\varv_\text{crit}^2=\frac{GM}{R_\text{e,crit}}\;,
	\end{equation}
	with $R_\text{eq}$ and $R_\text{pol}$ being the equatorial and the polar radii respectively, and $R_\text{e,crit}$ the [maximum] equatorial radius when $\varv_\text{crit}$ is reached.
	The standard setup has been $\Omega=0.4$ (either $\omega/\omega_\text{crit}$ or $\varv/\varv_\text{crit}$), based on the survey on Galactic BSG performed by \citet{huang10}.
	For the purpose of this paper, we set \text{for both \texttt{MESA} and \textsc{Genec} codes} the initial rotation as $\Omega=\omega/\omega_\text{crit}=0.4$ (equivalent to $\varv/\varv_\text{crit}\simeq0.276$ in \textsc{Genec}).

	The treatment of the stellar rotation also varies between both codes, not only for the mixing for also for the enhancement of the mass loss.
	\text{By default} \texttt{MESA} follows the standard correction factor from \text{\citet{friend86} and \citet{langer98}}
	\begin{equation}\label{corrmdot1}
		\dot M(\omega)=\dot M(\omega=0)\left(\frac{1}{1-\omega/\omega_\text{crit}}\right)^{0.43}\;,
	\end{equation}
	with the ratio $\omega/\omega_\text{crit}$ given by Eq.~\ref{omega_crit}.
	\texttt{MESA} also incorporates a limit to the maximum mass loss following \citet{yoon10}, to avoid large divergences of Eq.~\ref{corrmdot1} due to the dependence of $\omega_\text{crit}$ on the Eddington factor $\Gamma_\text{Edd}$.

	On the contrary, $\textsc{Genec}$ uses as correction factor the more complex formula provided by \citet{maeder00}
	\begin{equation}\label{corrmdot2}
		\dot M(\omega)=\dot M(0)\frac{(1-\Gamma_\text{e})^{\frac{1}{\alpha}-1}}{\left[1-\frac{\omega^2}{2\pi G\rho_m}-\Gamma_\text{e}\right]^{\frac{1}{\alpha}-1}}\;,
	\end{equation}
	with $\rho_m$ being the internal average density, $\alpha$ the CAK line-force parameter, and $\Gamma_\text{e}$ the electron scattering Eddington factor as given by Eq.~\ref{Gammae}.
	Eq.~\ref{corrmdot2} considers the break-up velocity not only with respect to the rotation but also with respect to the proximity to the Eddington limit ($\Omega\Gamma$-limit), together with being in rule with the so-called $\Omega$-slow solutions \citep{michel04,araya18,alex23a}.

	\begin{figure*}[t!]
		\centering
		\includegraphics[width=0.447\linewidth]{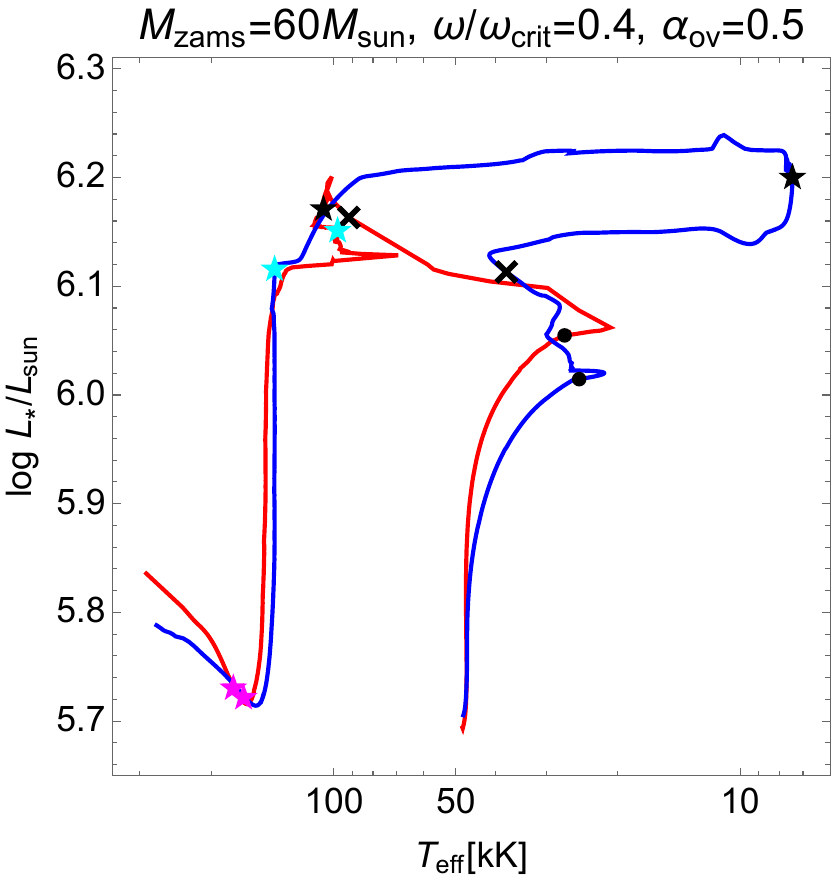}
		\hspace{1cm}
		\includegraphics[width=0.45\linewidth]{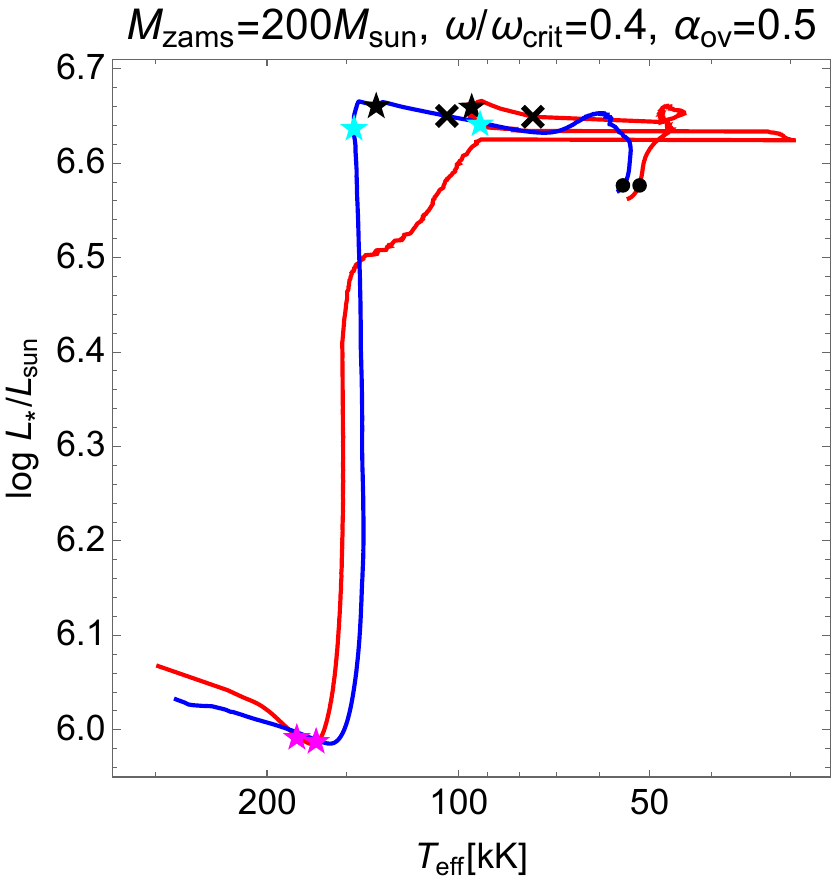}
		\caption{\small{HR diagram for \textsc{Genec} (red) and \texttt{MESA} (blue) evolution models with $60$ $M_\odot$ (left panel) and $200$ $M_\odot$ (right panel).
		Black dots represent the point where $\Gamma_\text{e}=0.5$ and thus the wind regime transits from thin to optically thick, black crosses represent the end of the H-core burning stage, black stars represent the beginning of the He-core burning stage, cyan stars represent the switch between H-rich to H-poor helium stars, and magenta stars represent the end of the He-core burning stage.}}
		\label{fullHRD}
	\end{figure*} 

%_____EVOLUTION THROUGH THE H-CORE BURNING STAGE__________________________________________
\section{Stellar evolution adopting new winds}\label{mainresults}
\subsection{H-core burning stage}
	\begin{table*}[t!]
		\centering
		\caption{Properties of our evolution models during the H-core burning stage.}
		\resizebox{\linewidth}{!}{
		\begin{tabular}{cc|ccccccccccccc}
			\hline
			\hline
			\multicolumn{2}{c}{Initial parameters} & \multicolumn{13}{c}{H-core burning}\\
			$M_\text{zams}$ & code & $\tau_\text{switch}$ & $M_\text{switch}$ & $X_\text{surf,switch}$ & $Y_\text{surf,switch}$ & X(N)$_\text{surf,switch}$ & $\tau_\text{MS}$ & $M_\text{final}$ & $M_\text{core}$ & $R_\text{max}$ & $X_\text{surf}$ & $Y_\text{surf}$ & X(N)$_\text{surf}$ & $\Gamma_\text{e}$\\
			$[M_\odot]$ & & [Myr] & $[M_\odot]$ & \multicolumn{3}{c}{mass fraction} & [Myr] & $[M_\odot]$ & $[M_\odot]$ & $[R_\odot]$ & \multicolumn{3}{c}{mass fraction} &\\
			\hline
			60 &  \textsc{Genec} & 3.959 & 53.6 & 0.514 & 0.472 & 0.008 & 4.284 & 42.1 & 33.6 & 82.6  & 0.057 & 0.929 & 0.008 & 0.562\\
			60 &  \texttt{MESA} & 3.734 & 51.9 & 0.632 & 0.354 & 0.007 & 4.186 & 44.5 & 37.2 & 73.1 & 0.417 & 0.569 & 0.009 & 0.605\\
			\hdashline
			200 & \textsc{Genec} & 0.183 & 198.9 & 0.717 & 0.269 & 0.002 & 2.517 & 100.4 & 93.3 & 36.4 & 0.013 & 0.973 & 0.008 & 0.691\\
			200 & \texttt{MESA} & 0.092 & 199.3 & 0.717 & 0.268 & 8$\times10^{-4}$ & 2.543 & 100.7 & 100.7 & 23.6 & 0.021 & 0.966 & 0.009 & 0.696\\
			\hline
		\end{tabular}}
		\label{table_finalmodels}
	\end{table*}
	\begin{figure*}[t!]
		\centering
		\includegraphics[width=0.24\linewidth,align=c]{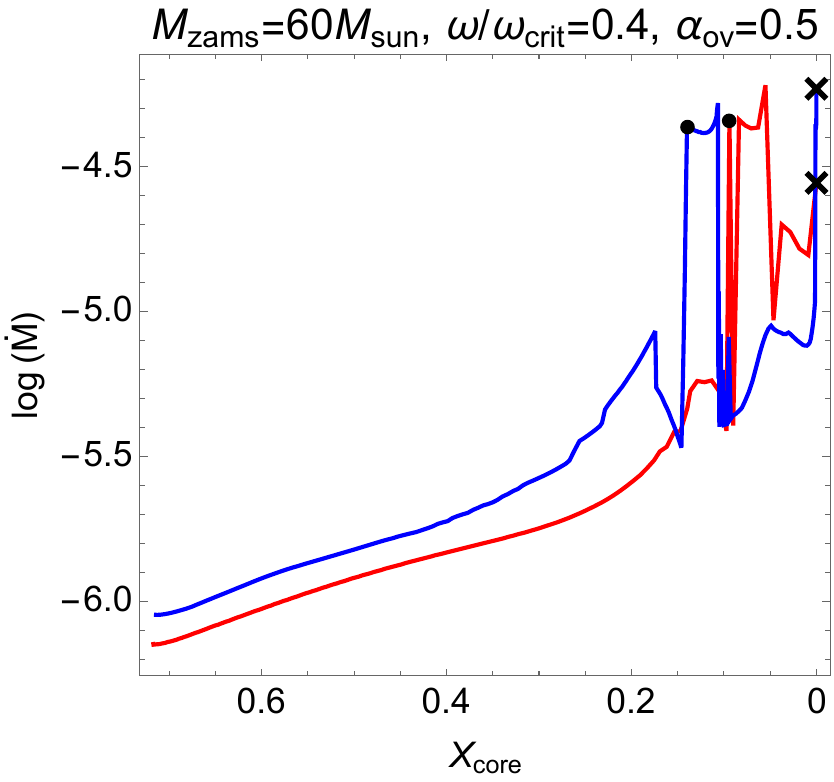}
		\includegraphics[width=0.24\linewidth,align=c]{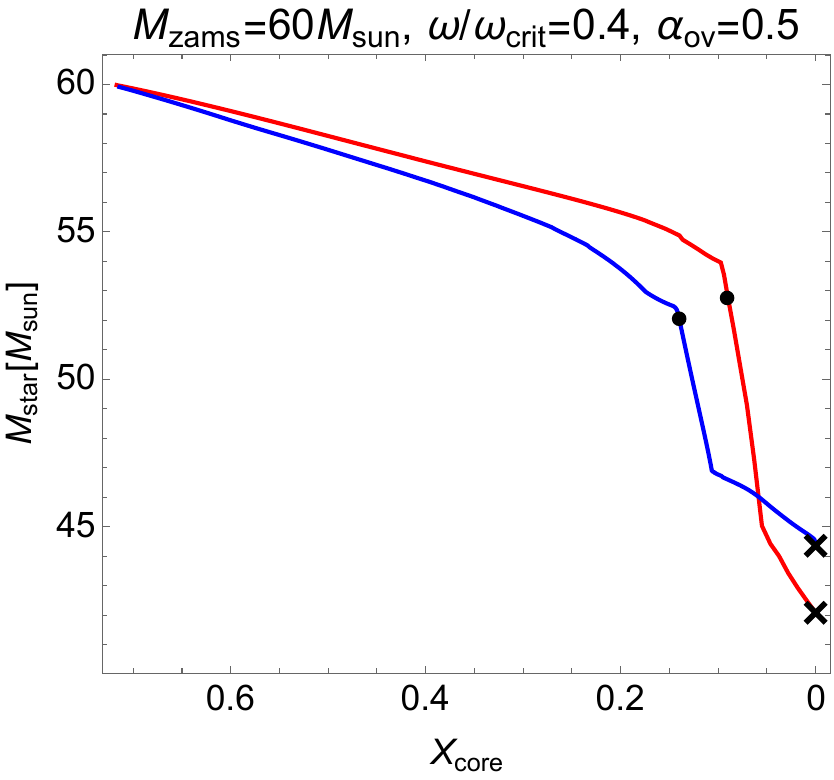}
		\includegraphics[width=0.24\linewidth,align=c]{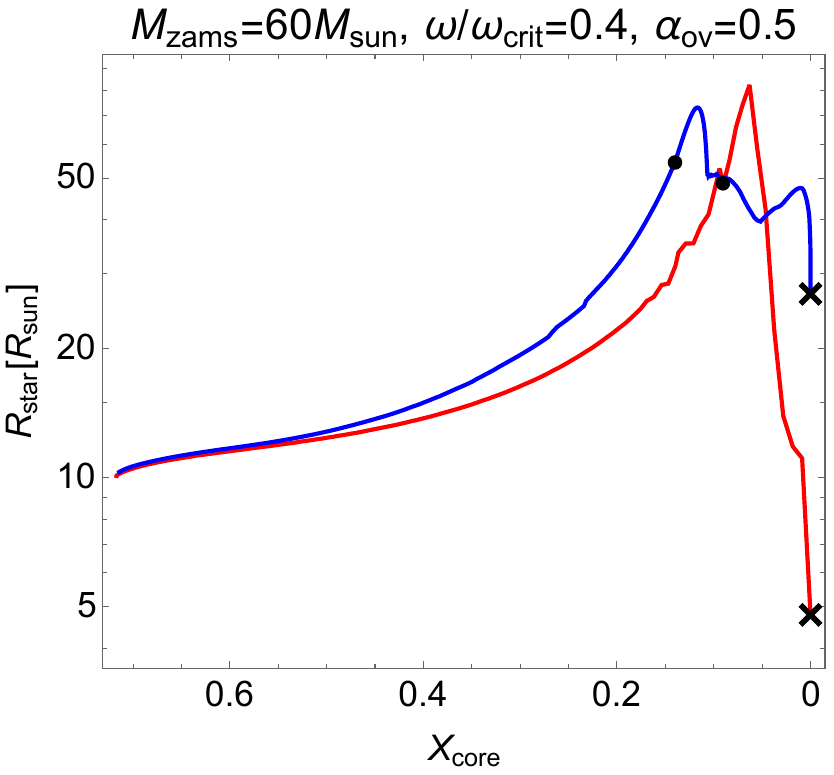}
		\includegraphics[width=0.24\linewidth,align=c]{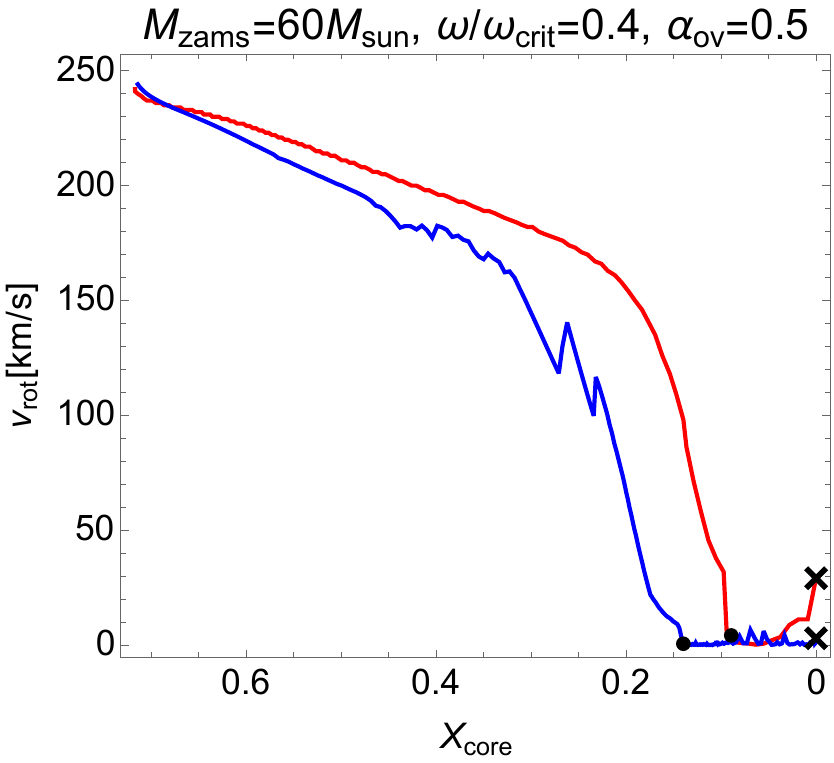}\\
		\includegraphics[width=0.24\linewidth,align=c]{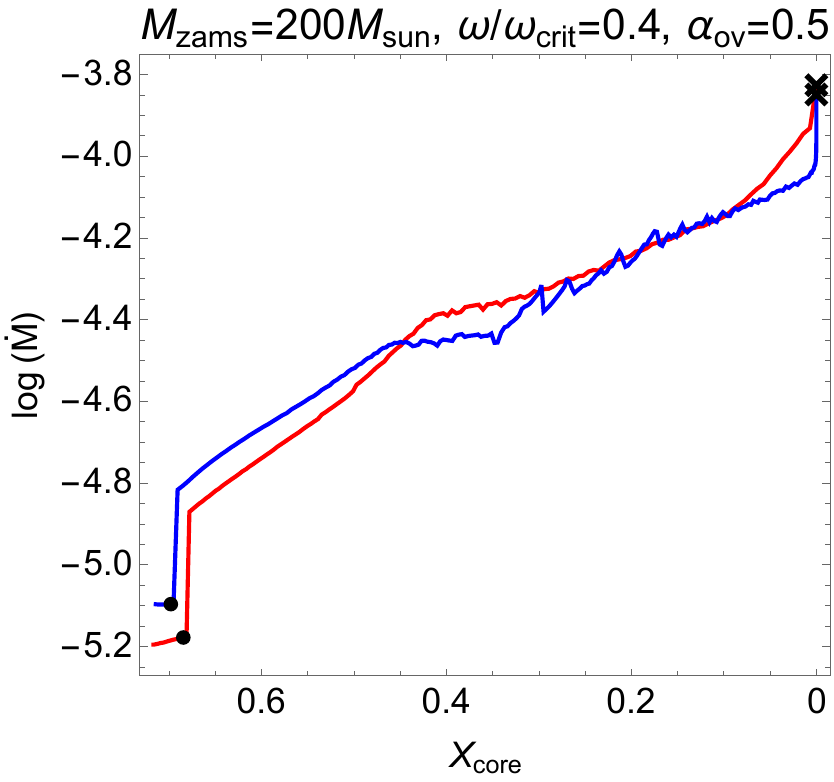}
		\includegraphics[width=0.24\linewidth,align=c]{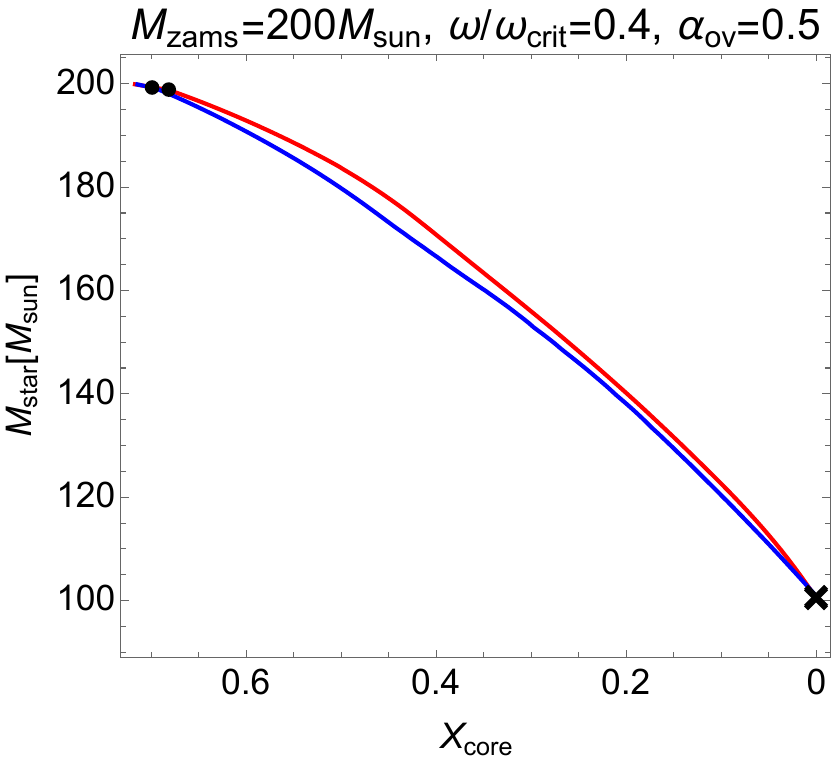}
		\includegraphics[width=0.24\linewidth,align=c]{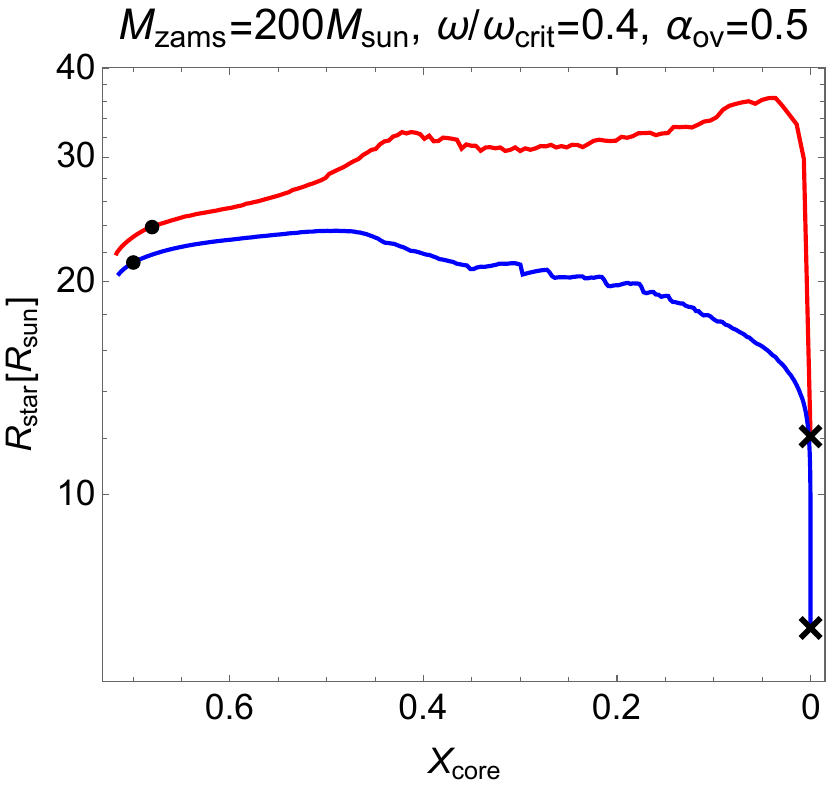}
		\includegraphics[width=0.24\linewidth,align=c]{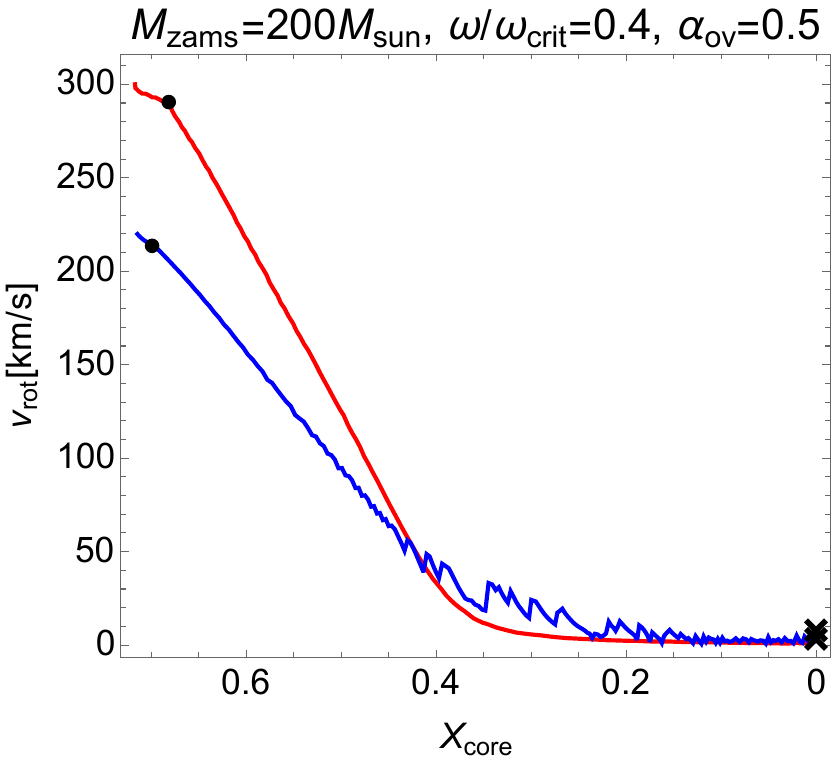}
		\caption{\small{Left panel: Evolution of mass-loss rate, stellar mass, stellar radius, and Eddington factor during the H-core burning stage for \textsc{Genec} and \texttt{MESA} evolution models with $M_\text{zams}=60\,M_\odot$.
		Right panel: same, but for evolution models with $M_\text{zams}=200\,M_\odot$.}}
		\label{plots_hydrogenburning}
	\end{figure*}

	Fig.~\ref{fullHRD} shows the evolutionary tracks across the HR diagram according to our new \textsc{Genec} and \texttt{MESA} models, for both $60\,M_\odot$ and $200\,M_\odot$.
	Fig.~\ref{plots_hydrogenburning} shows the evolution of the mass loss, the stellar mass, the stellar radius, and the rotational velocity, from the ZAMS until the end of the H-core burning stage.

	We observe that, even though the ZAMS point in the HRD is almost the same for the \texttt{MESA} and \textsc{Genec} tracks, \texttt{MESA} models start with a higher mass loss due to the more rigid effect from Eq.~\ref{corrmdot1}, making a difference of $\sim0.2$ dex in logarithmic scale.
	However, for our $M_\text{zams}=60$ $M_\odot$ models, this difference is not large enough to produce significant deviations at the beginning, and thus both tracks move upwards increasing luminosity at temperature nearly constant.
	Both models reach $\Gamma_\text{e}=0.5$ prior of reaching their maximum radial expansion in the MS, but our \textsc{Genec} model reaches a higher temperature because of its slightly slower mass loss, its faster rotation, and a more efficient mixing.
	This is evidenced in Table~\ref{table_finalmodels}, where the \texttt{MESA} model reaches to the switch point from thin to thick winds (black dots in Figures~\ref{fullHRD} and \ref{plots_hydrogenburning}) with lower mass and lower mass fractions of helium and nitrogen, in contrast with the \textsc{Genec} model.
	Once the winds for both models become optically thick, the mass loss in both cases is described either by Eq.~\ref{mdotthick} or V01/GH08 formulae depending on the condition of $T_\text{eff}$ being higher or lower than 30 kK.
	Because our \textsc{Genec} model has a larger luminosity and larger $\Gamma_\text{e}$, its optically thick $\dot M$ will be larger enough to remove more layers and therefore to end the H-core burning stage with less mass than our \texttt{MESA} model ($42.1\,M_\odot$ vs $45\,M_\odot$).
	However, the most remarkable difference is the final chemical composition at the H-depletion: our \textsc{Genec} model ends with a helium mass fraction of $Y_\text{surf}=0.929$, i.e., stellar structure is almost homogeneous, and much more compact ($\simeq5\,R_\odot$) and hotter ($T_\text{eff}\simeq90$ kK).
	
	Notice that, for both $60\,M_\odot$ models, the transition occurs outside the range of validity of both $\dot M_\text{thin}$ and $\dot M_\text{thick}$, where V01 dominates both wind regimes \citep[Fig.~1]{romagnolo24} and the only jump in mass loss is due to the bi-stability jump \citep{vink99}, at $T_\text{eff}\simeq28$ kK.
	Despite this, the fact of reaching $\Gamma_\text{e}=0.5$ during the main sequence implies that this point also represents the transition from thin to thick winds, as we will discuss later in Sec.~\ref{selfconsistent_gamma05}.

	For our $M_\text{zams}=200$ $M_\odot$ models, the initial mass loss start around $10^{-5}M_\odot$ yr$^{-1}$, and again the \texttt{MESA} model starts with a higher $\dot M$ due to Eq.~\ref{corrmdot1}.
	The initial rotational velocity substantially differs between both models, because both models calculate a slightly different starting point for the ZAMS in the HRD as a consequence of \text{both} the numerical uncertainties \citep[][which tend to increase for larger masses]{agrawal22}, and the different calculation of the critical rotational velocity outlined in Sec.~\ref{rotationmethod}.
	Despite these corrections, both models \text{show a very similar track across the HRD, reaching} $\Gamma_\text{e}=0.5$ at the very beginning of their lives and subsequently the stars spend more than their $90\%$ of their H-core burning lifetime in the optically thick wind regime.
	Nonetheless, in this mass regime the mass-loss escalates with metallicity only as $\dot M\propto Z^{0.48}$ (according to Eq.~\ref{mdotthick}) and thus our initial mass loss for the thick wind regime is around $\log\dot M\simeq-4.8$, not large enough to produce an abrupt decrease in the luminosity, reason why both tracks keep moving upwards in the HRD diagram until $\log L/L_\odot\simeq6.7$.
	Once the mass loss increases enough to stop the rise in luminosity, both models move bluewards keeping $L_*$ almost constant.
	This time, the \texttt{MESA} model drops its rotational velocity slower than the \textsc{Genec} model, keeping a more moderate radius and ending their H-core burning stage with a larger effective temperature ($\sim100$ kK, against the $\sim80$ kK from the \textsc{Genec} model).
	Despite that our \textsc{Genec} expanded more redwards, it ends its H-burning with a more advanced chemical enrichment at the surface.
	Even though this could be partially attributed to the larger initial rotation, it is also because that \textsc{Genec} implements a more efficient rotational mixing \citep{meynet00,choi16}.

	The correlation between the both tracks shows the most important difference with respect to previous evolution studies adopting V01, where it is found that for $M_\text{zams}=200$ $M_\odot$ the mass loss starts being strong enough to make the star quickly drop in luminosity \citep{yusof13,sabhahit22,sabhahit23,martinet23}.
	Differences in mass loss also carries another important discrepancy with respect previous studies: our models end the H-core burning stage with final masses of $\sim100.5$ $M_\odot$.
	This is larger than the $M_\text{final}\simeq85\,M_\odot$ found by \citet{martinet23} for $M_\text{zams}=180\,M_\odot$, which is the study with the structure setup closest to ours ($\alpha_\text{ov}=0.2$ and Ledoux convection criterion) but still adopting old winds; but our result is even more remarkably larger than the $M_\text{final}\simeq40\,M_\odot$ found by \citet{sabhahit22,sabhahit23}, who adopted a $\dot M_\text{thick}$ from \citet{vink11}.
	In contrast with these huge mass loss, our models adopting weaker (thin and thick) winds makes a $M_\text{zams}=200\,M_\odot$ star lose `only' one half of their mass at the end of H-core burning stage.
	This carries an important impact related to the spectroscopic diagnostics to be done over Galactic VMS, as we will see in Sec.\ref{spectroscopicanalysis}.

%_____EVOLUTION THROUGH THE He-CORE BURNING STAGE__________________________________________
\subsection{He-core and C-core burning stages}\label{heliumcoreevolution}
	\begin{table*}[t!]
		\centering
		\caption{Properties of our evolution models during the He-core and C-core burning stages.}
		\resizebox{\linewidth}{!}{
		\begin{tabular}{cc|cccccccc|cccccccc}
			\hline
			\hline
			\multicolumn{2}{c}{Initial parameters} & \multicolumn{8}{c}{He-core burning} & \multicolumn{7}{c}{C-core burning}\\
			$M_\text{zams}$ & code & $\tau_\text{He}$ & $M_\text{final}$ & $M_\text{core}$ & $R_\text{max}$ & $Y_\text{surf}$ & X(C)$_\text{surf}$ & X(O)$_\text{surf}$ & $\Gamma_\text{e}$ & $\tau_\text{C}$ & $M_\text{final}$ & $R_\text{max}$ & $Y_\text{surf}$ & X(C)$_\text{surf}$ & X(O)$_\text{surf}$ & $\Gamma_\text{e}$\\
			$[M_\odot]$ & & [Myr] & $[M_\odot]$ & $[M_\odot]$ & $[R_\odot]$ & \multicolumn{3}{c}{mass fraction} & & [Myr] & $[M_\odot]$ & $[R_\odot]$ & \multicolumn{3}{c}{mass fraction}\\
			\hline
			60 & \textsc{Genec} & 4.657 & 16.7 & 11.9 & 7.1 & 0.059 & 0.317 & 0.600 & 0.497 & 4.662 & 16.1 & 0.79 & 0.044 & 0.286 & 0.646 & 0.654\\
			60 & \texttt{MESA} & 4.546 & 16.6 & 16.6 & 763.0 & 0.059 & 0.345 & 0.578 & 0.482 & 4.552 & 16.1 & 0.95 & 0.047 & 0.319 & 0.616 & 0.589\\
			\hdashline
			200 & \textsc{Genec} & 2.835 & 25.4 & 18.4 & 12.1 & 0.032 & 0.200 & 0.746 & 0.596 & 2.839 & 24.4 & 1.0 & 0.020 & 0.171 &  0.784 & 0.736\\
			200 & \texttt{MESA} & 2.855 & 25.7 & 25.7 & 11.6 & 0.036 & 0.230 & 0.714 & 0.578 & 2.860 & 24.8 & 1.27 & 0.026 & 0.205 & 0.748 & 0.668\\
			\hline
		\end{tabular}}
		\label{table_HeCfinalmodels}
	\end{table*}
	\begin{figure*}[t!]
		\centering
		\includegraphics[width=0.24\linewidth,align=c]{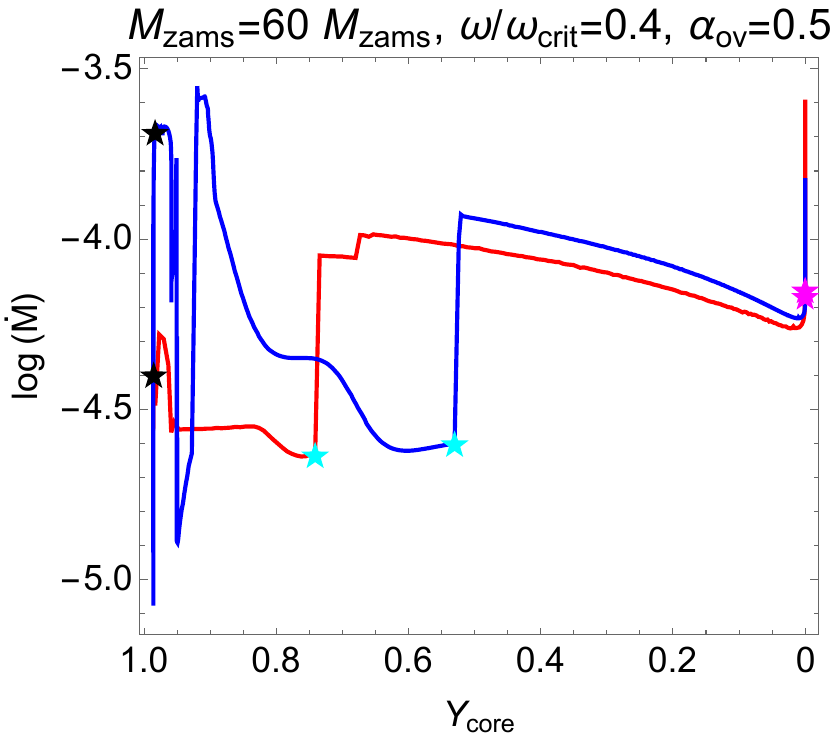}
		\hspace{1cm}
		\includegraphics[width=0.24\linewidth,align=c]{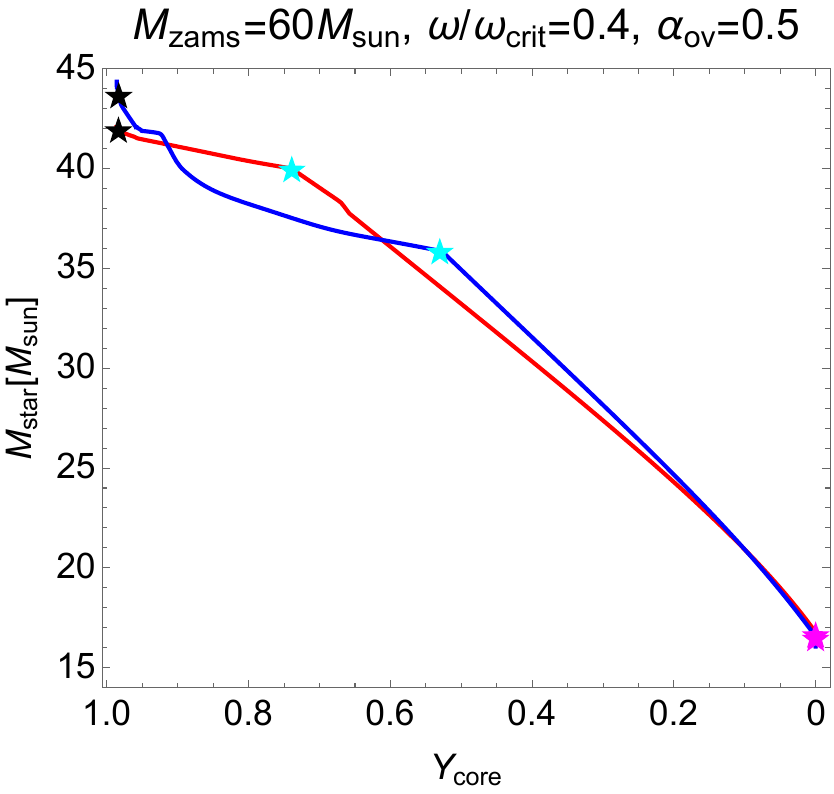}
		\hspace{1cm}
		\includegraphics[width=0.24\linewidth,align=c]{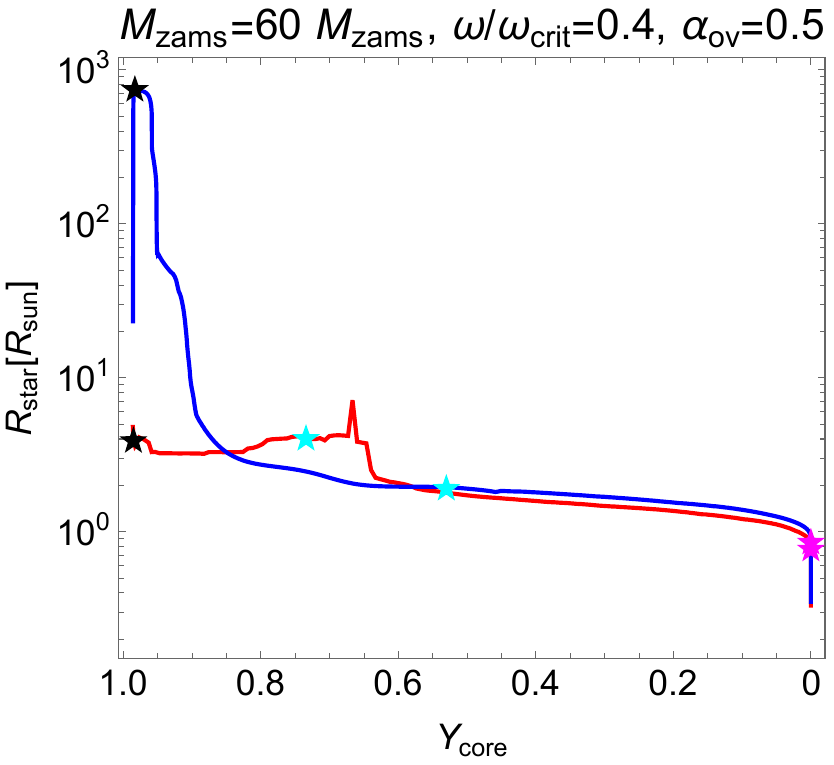}\\
		\includegraphics[width=0.24\linewidth,align=c]{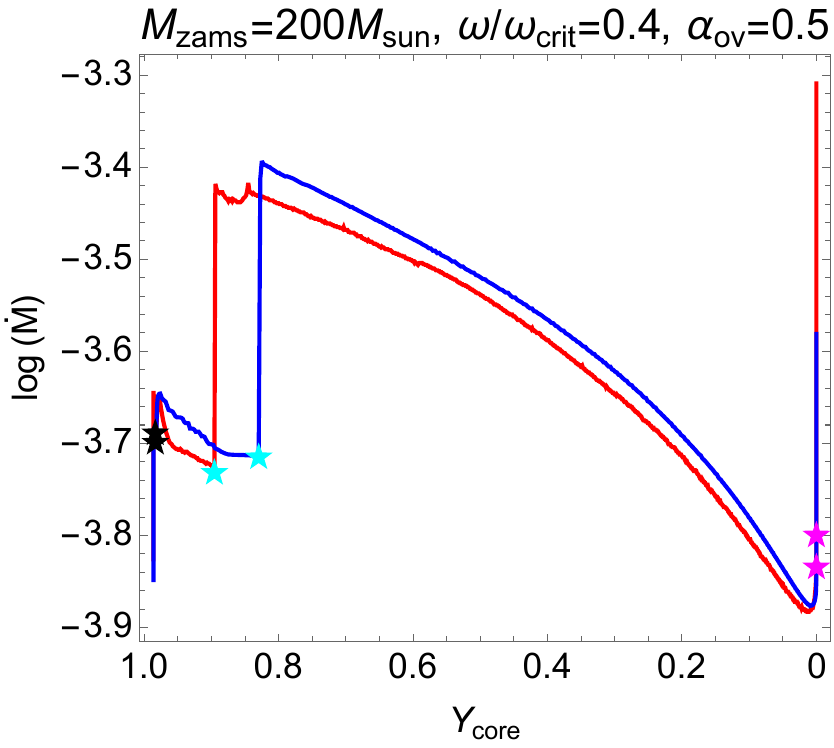}
		\hspace{1cm}
		\includegraphics[width=0.24\linewidth,align=c]{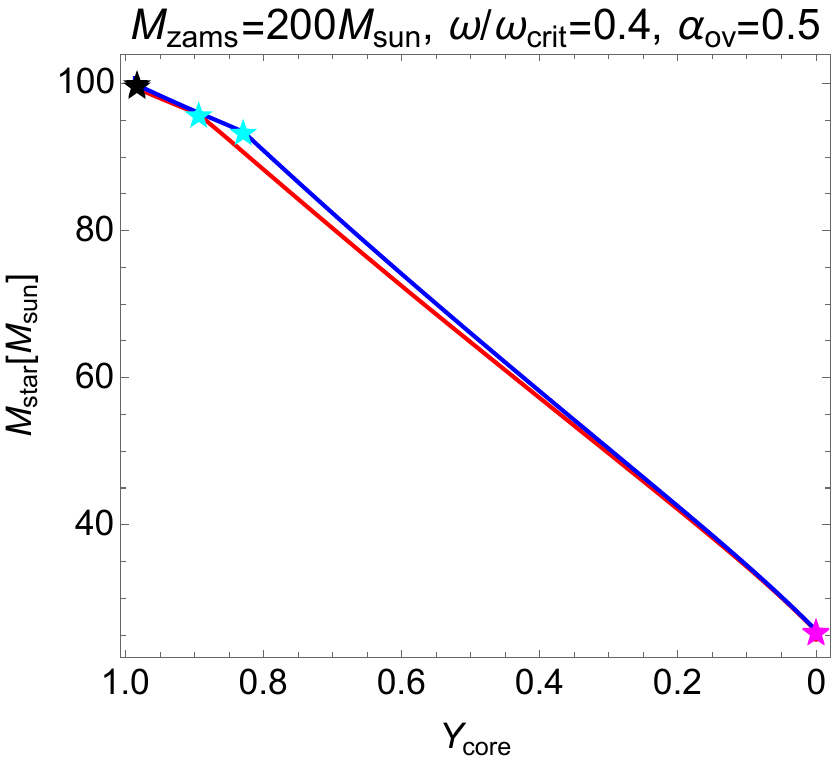}
		\hspace{1cm}
		\includegraphics[width=0.24\linewidth,align=c]{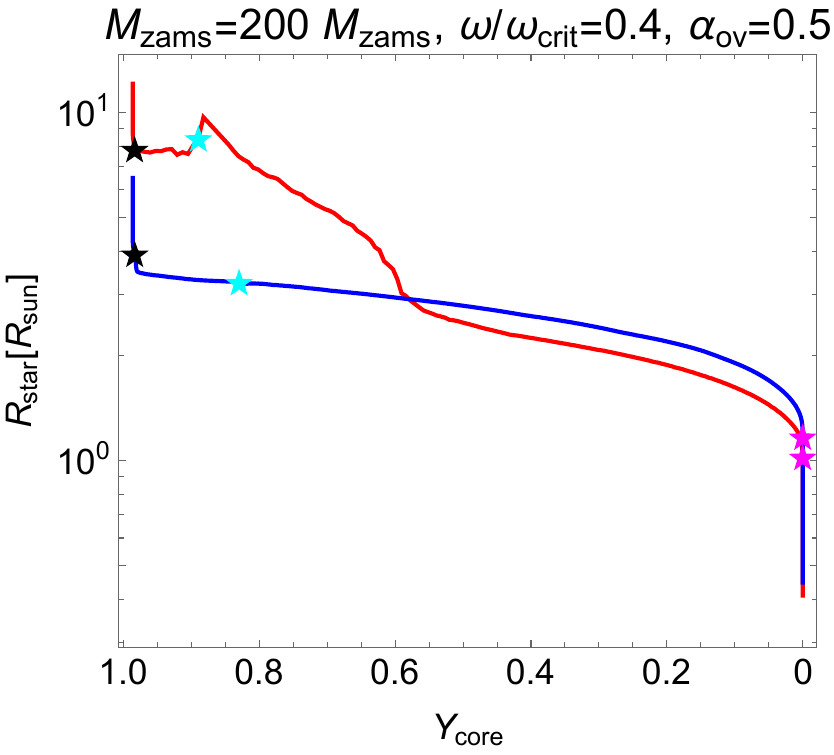}
		\caption{\small{Left panel: Evolution of mass-loss rate, stellar mass, stellar radius, and Eddington factor during the He-core and C-core burning stages for \textsc{Genec} and \texttt{MESA} evolution models with $M_\text{zams}=60\,M_\odot$.
		Right panel: same, but for evolution models with $M_\text{zams}=200\,M_\odot$.}}
		\label{plots_heliumcarbonburning}
	\end{figure*}

	Given that the star $60\,M_\odot$ from our \texttt{MESA} model ends the MS with a less homogeneous chemical structure (only a $\sim57\%$ of helium mass fraction, in contrast with the $\sim93\%$ predicted by \textsc{Genec}), it quickly expands until reaching $T_\text{eff}\simeq8\,000$ K with a maximum radius of $\simeq740\,R_\odot$.
	This is because stars keeping a more constant chemical composition from the core to the surface remain in the bluer part of the HR diagram \citep{maeder87}.
	The beginning of the He-core burning stage (black star symbol in the Figures) makes the star return bluewards, thus increasing its temperature and its mass loss.
	On the contrary, \textsc{Genec} model predicts a compact star which enters into the He-core burning stage without any radial expansion.
	This divergence on the models indicate that $M_\text{zams}\simeq60\,M_\odot$ represents a threshold for the Hertzsprung-gap: less massive stars will all expand post MS despite minor differences in the rotational mixing, whereas more massive will not \citep[see Fig.~2 from][]{romagnolo24}.
	These initial differences during He-core ignition later converge into the position at the HRD where both stellar models become H-poor WR stars.
	Hence, both models show a similar track for the rest of the He-core burning, with a large diminishing in the luminosity from $\log L/L_\odot\simeq6.2$ to $\simeq5.75$ at temperature almost constant $T_*\simeq140$ kK.
	As a consequence, the final masses at the end of the He-core burning stage are relatively the same for \texttt{MESA} and \textsc{Genec} tracks ($\simeq16.7\,M_\odot$), and thus the remaining evolution through the C-core burning is practically identical for both models.

	For the $200\,M_\odot$ case, differences in both the stellar mass and the chemical composition after the H-depletion are almost negligible, and just the HRD where the He-ignition occurs is significantly relevant.
	Because of its proximity to the Eddington limit (with $\Gamma_\text{e}\simeq0.7$) and lower density, the convection at the outer envelope of a star born with $M_\text{zams}=200\,M_\odot$ is highly superadiabatic (i.e, the temperature gradient is larger than the adiabatic temperature gradient) and therefore the energy released in the interior remains locked at the subsurface layers, thus creating numerical instabilities in the models where each evolution code adopts different strategies to handle it \citep{agrawal22b}. 
	\texttt{MESA} reduces the superadiabacity in radiatively-dominated regions by means of the \texttt{use\_superad\_reduction} method, which can be interpreted as a lowering in the opacity $\kappa(r)$ in the sub-surface regions near the Eddington limit \citep{jermyn23}.
	On the contrary, our 200 $M_\odot$ \textsc{Genec} does not incorporate this superadiabacity rescaling and thus excess in the temperature gradient leads into a considerable inflation when the last traces of hydrogen are removed from the surface and the opacity drastically changes \citep{yusof13,martinet23}, until this excess of energy is released due to Supra-Eddington mass loss \citep{ekstrom12} and the star returns to its initial radius.
	During this expansion, only less than the $0.1\%$ of the total mass is inflated, keeping the remaining $99.9\%$ of the total mass constrained into the same stellar radius, reason why this expansion should not be considered for any binary interaction \cite{belczynski22a}.
	Indeed, we show this expansion due to superadiabatic instability as a sudden expansion redwards in the HRD of Fig.~\ref{fullHRD}, but we do not consider it for our plot of radial evolution during the He-core burning (Fig.~\ref{plots_heliumcarbonburning}) nor for the maximum radius in Table~\ref{table_HeCfinalmodels}.
	Moreover, the strategies to deal with these superadiabatic instabilities are approached from a numerical point of view, and possibilities of a physical phenomenon have been barely considered.
	After this phenomenon, the star will decrease its luminosity at almost constant temperature until $\log L/L_*\simeq6.0$ according to both \texttt{MESA} and \textsc{Genec} models, then the He-core burning stage ends with a stellar mass of $\simeq25.5\,M_\odot$ being the star a ball of $\simeq94\%$ carbon and oxygen.
	We ultimately reach the end of the C-core burning stage with final masses of $24.4\,M_\odot$ according to \textsc{Genec} and $24.8\,M_\odot$ according to \texttt{MESA}.

%_____SPECTROSCOPIC ANALYSIS
	\begin{table*}[t!]
		\centering
		\caption{Lifetime (in Myr) predicted by our \textsc{Genec} and {\tt MESA} models in the spectroscopic phases according to stellar evolution.
		The percentages with respect to the total lifetime are shown in parenthesis.}
		\resizebox{\linewidth}{!}{
		\begin{tabular}{cc|cccccccccccccc}
			\hline
			\hline
			$M_\text{zams}$ & code & $\Delta\tau_\text{O}$ & $\Delta\tau_\text{B}$ & $\Delta\tau_\text{WNL,$X_\text{surf}$>0.3}$ & $\Delta\tau_\text{LBV}$ & $\Delta\tau_\text{WNL,$X_\text{surf}$<0.3}$ & $\Delta\tau_\text{WNE}$ & $\Delta\tau_\text{WC}$ & $\Delta\tau_\text{WO}$\\
			\hline
			60 &  \textsc{Genec} & $3.942\,(84.55\%)$ & $0.017\,(0.38\%)$ & 0.205\,(4.40\%) & 0\,(0\%) & 0.141\,(3.01\%) & 0.069\,(1.48\%) & 0.283\,(6.08\%) & 0.005\,(0.10\%)\\
			60 &  \texttt{MESA} & 3.701\,(81.29\%) & 0.031\,(0.67\%) & 0.461\,(10.14\%) & 0.012\,(0.26\%) & 0.071\,(1.57\%) & 0.061\,(1.35\%) & 0.211\,(4.64\%) & 0.004\,(0.08\%)\\
			\hdashline
			200 & \textsc{Genec} & 0.183\,(6.45\%) & 0\,(0\%) & 1.679\,(59.14\%) & 0.006\,(0.22\%) & 0.574\,(20.21\%) &  0.108\,(3.79\%) & 0.285\,(10.03\%) & 0.004\,(0.15\%)\\
			200 & \texttt{MESA} & 0.092\,(3.21\%) & 0\,(0\%) & 1.794\,(62.72\%) & 0\,(0\%) & 0.602\,(21.04\%) & 0.096\,(3.37\%) & 0.273\,(9.55\%) & 0.003\,(0.11\%)\\
			\hline
		\end{tabular}}
		\label{table_spectroscopicphases}
	\end{table*}
\section{Theoretical and observational diagnostics}\label{diagnostics}

\subsection{Self-consistent mass loss close to $\Gamma_\text{e,trans}$}\label{selfconsistent_gamma05}
	As shown in Sec.~\ref{eddingtonlimit}, our evolution models use as a criterion to switch from optically thin winds of OB-type stars to thick winds of WR-type stars during the main sequence, the Eddington factor instead of the surface hydrogen abundance.
	In particular we use the transition defined by \citet{bestenlehner20}, which corresponds to the $\Gamma_\text{e}$ where the line-driven contribution to the CAK mass-loss equalises with the continuum-driven contribution.
	Below $\Gamma_\text{e,trans}$, the wind is predominantly line-driven and thus $\dot M$ it can be described by our self-consistent m-CAK prescription, whereas over $\Gamma_\text{e,trans}$ the continuum radiation by electron scattering dominates.
  	
\begin{figure}[t!]
		\centering
		\includegraphics[width=\linewidth]{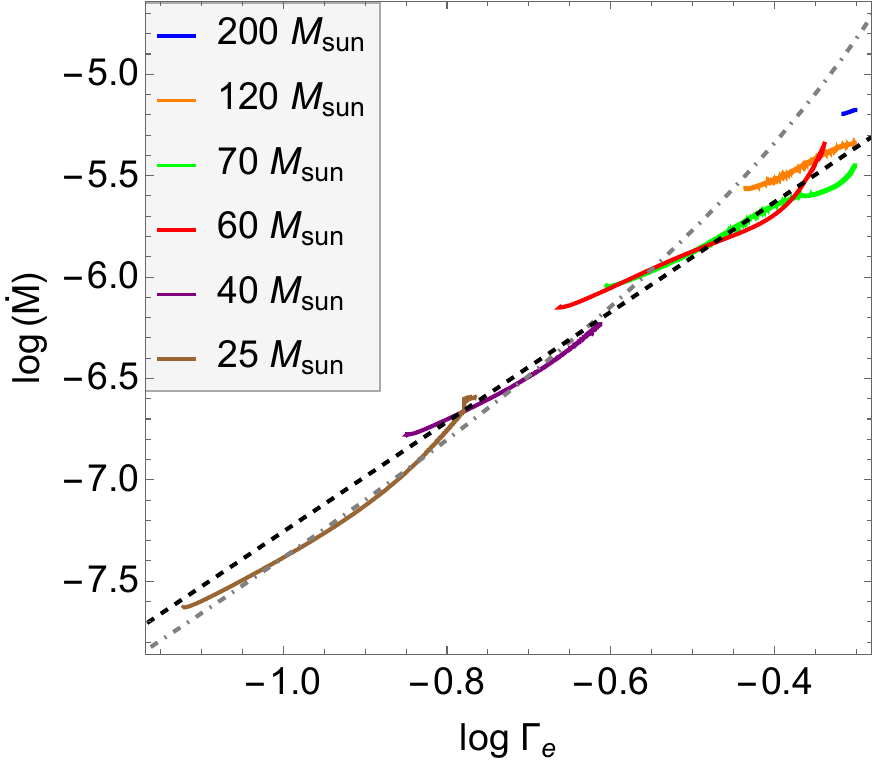}
		\caption{\small{$\dot M-\Gamma_\text{e}$ correlations taken from the \textsc{Genec} self-consistent evolution models from \citet{alex23a} and this work.
		The dashed line represents the best linear fit for these $M_\text{sc}$, whereas the gray dot-dashed line represents the fit of Eq.~\ref{mdotthick} adopted for $Z=0.014$.}}
		\label{mdot_vs_Gammae}
	\end{figure}

	Therefore, a question that arises is how the transition towards thick winds approaches from self-consistent mass-loss rates at optically thin regime towards $\Gamma_\text{e,trans}$.
	To do so, we plot in Fig.\ref{mdot_vs_Gammae} the $\dot M_\text{sc}$ as a function of the Eddington factor coming from the models introduced in \citet{alex23a}, plus the \textsc{Genec} models introduced in this work.
	We can see that the $\dot M-\Gamma_\text{e}$ easily correlates with a linear fit in the log-log plane, as
	\begin{equation}\label{MdotGammathin}
		\log \dot M_\text{sc}\simeq-4.547+2.709\log\Gamma_\text{e}\;
	\end{equation}
	this rough fit of Eq.~\ref{mdotalex22bnew} agrees reasonably well with the $\dot M\propto\Gamma_\text{e}^{2.73\pm0.43}$ from \citet{bestenlehner14} and with the $\dot M\propto\Gamma_\text{e}^{2.69}$ from \citep{brands22}, for the low Eddington factor regime.
	In other words, $\dot M_\text{sc}$ behaves like Eq.~\ref{mdotthick} when $\Gamma_\text{e}\lesssim0.5$, which is appreciated in Fig.~\ref{mdot_vs_Gammae}, where the both fits run in parallel for $\Gamma_\text{e}\ll1$.
	It is important to mention that even though the $\dot M-\Gamma_\text{e}$ correlation is a very useful tool to describe the escalation of the mass loss in the proximity of the Eddington limit, it is not a formal function.
	The value of $\dot M$ may vary for a same value of $\Gamma_\text{e}$, if the stellar mass or the luminosity or the hydrogen abundance are different.
	In particular, the linear behaviour of $\dot M_\text{sc}$ dissipates as we approach $\Gamma_\text{e}\simeq0.5$, thus suggesting that the m-CAK prescription stops being valid for regimes with larger Eddington factor (where the contribution of the electron scattering to the radiative acceleration becomes more dominant) and a prescription considering optically thick winds needs to be adopted.
	
	We also observe in Fig.~\ref{mdot_vs_Gammae} that Eq.~\ref{MdotGammathin} and Eq.~\ref{mdotthick} intersect at $\log\Gamma_\text{e}\simeq-0.62$ ($\Gamma_\text{e}\simeq0.24$), very close to the empirical kink found by \citet{bestenlehner14} at $\log\Gamma_\text{e}\simeq-0.58$.
	After that point, formula for $\dot M_\text{thick}$ winds clearly exceeds $\dot M_\text{thin}$, finishing with a difference of $\simeq0.5$ dex when $\Gamma_\text{e}=0.5$, as shown in Fig.~\ref{plots_hydrogenburning}.
	Hence, we could have a smooth transition from $\dot M_\text{thin}$ to $\dot M_\text{thick}$ by simply setting the transition at the point where both formulae intersect.
	However, we decide for the purposes of this paper to employ $\Gamma_\text{e,trans}$ as originally set by \citet{bestenlehner20}, to preserve the physical meaning and exploring the caveats.

	The rough fit introduced in Eq.~\ref{MdotGammathin} is set for solar metallicity, whereas the fit introduced in Eq.~\ref{mdotthick} corresponds to the calibration done by \citet{brands22} at LMC metallicity ($Z=0.006$) plus an extra metallicity-dependence prescribed by us.
	Actually, the general expression derived by \citet{bestenlehner20} is
	\begin{equation}\label{Best20generalMdot}
		\log\dot M=\log M_0+\left(\frac{1}{\alpha}+0.5\right)\log(\Gamma_\text{e})-\left(\frac{1-\alpha}{\alpha}+2\right)\log(1-\Gamma_\text{e})\;,
	\end{equation}
	with $\alpha$ being the CAK line-force parameter and $\dot M_0$ the mass-loss rate at $\Gamma_\text{e,trans}$.
	In principle, we could merge both Eq.~\ref{MdotGammathin} and Eq.~\ref{mdotthick} into one single expression as Eq.~\ref{Best20generalMdot}, by using the tabulated $\alpha$ values from \citet{alex23a}.
	In such case, the transition from thin to thick winds would be smooth, given that the continuum already contributes to the wind mass-loss rate even below $\Gamma_\text{e,trans}=0.5$.
	However, that merging would require the analysis of different metallicities, and the present study is focused on the Milky Way.
	In a forthcoming paper we aim to explore the possibility of adopting one unique Eq.~\ref{Best20generalMdot} for both thin and thick winds, together with more self-consistent calculations of the line-force parameters \citep{alex22a} for stellar models around $\Gamma_\text{e,trans}$.

% Eddington factor and wind efficiency
\subsection{Eddington factor and wind efficiency}\label{comparisonwitheta}
	The transition from thin to thick winds based on the proximity of the Eddington factor has been recently studied by \citet{grafener21} and \citet{sabhahit23}, where the switch from OB to WNh stars is set by using the relationship between the optical depth at the sonic point of the wind structure ($\tau_\text{s}$) and the wind efficiency number ($\eta\equiv\dot M\varv_\infty/(L_*/c)$) from \citet{grafener17}
	\begin{equation}\label{opticaldepth}
		\tau_\text{s}\simeq\frac{\dot M \varv_\infty}{L_*/c}\left(1+\frac{\varv_\text{esc}^2}{\varv_\infty^2}\right)\;,
	\end{equation}
	with $\varv_\infty$ and $\varv_\text{esc}$ being the wind terminal and the escape velocities respectively.

	Since $\eta$ denotes the ratio between the mechanical wind momentum and the total momentum of the radiatively-driven wind, the value $\eta=1$ corresponds to the single scattering limit.
	If $\eta\ge1$, it means that photons can be absorbed and re-emitted through the wind structure more than once \citep{lucy93}.
	Performed calculations found $\eta\simeq0.6$ as the transition point for $Z_\odot$, based on the mass-loss rates for Galactic and LMC WNh stars \citep{vink12,sabhahit22}, plus finding that this $\eta_\text{switch}$ decreases for lower metallicities \citep{sabhahit23}.
	These calibrations make the value of this transition mass-loss rate $\dot M_\text{trans}$ being model-independent of factors such as clumping.
	However, the temporal location of this thin-to-thick transition depends on the evolution of $\eta$, which in turns depends on the mass-loss recipe adopted a priori for the initial optically thin regime (V01 for this case, even for $M_\text{zams}\simeq40-60\,M_\odot$).
	\begin{figure}[t!]
		\centering
		\includegraphics[width=0.9\linewidth]{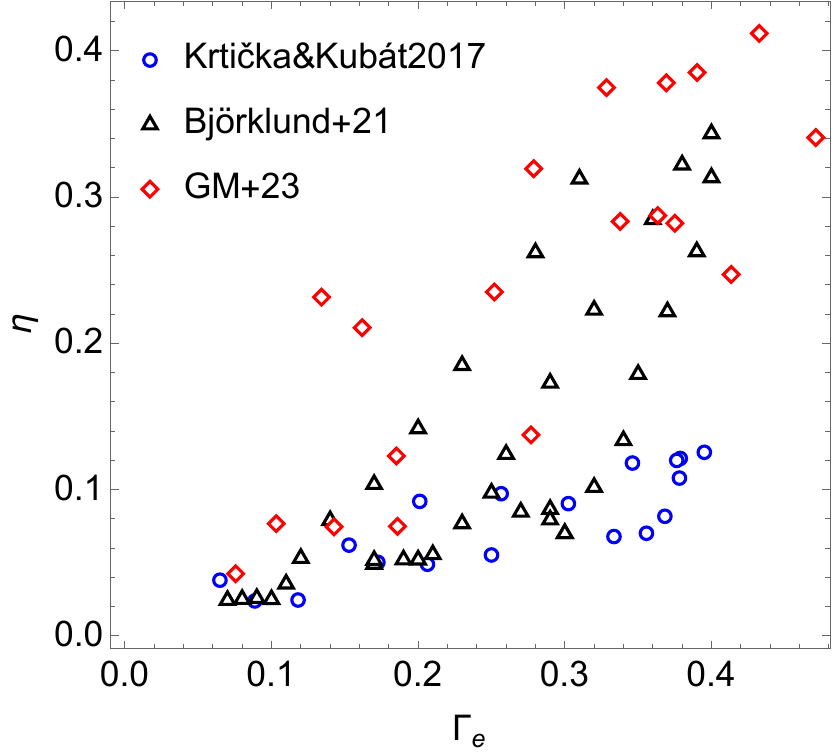}
		\caption{\small{$\eta-\Gamma_\text{e}$ correlations taken from theoretical models tabulated by \citet{kk17}, \citet{bjorklund21}, and \citet{alex23a}.}}
		\label{eta_vs_Gammae}
	\end{figure}
	
	On the contrary, the proposed transition towards thicker winds at $\Gamma_\text{e}=\kappa_\text{e}L_*/4\pi cGM_*=0.5$ is explicitly independent on the mass-loss regime adopted for the optically thin regime, together with not needing any $\beta$-law assumption for the velocity profile \citep{alex22a}.
	In this case, we enter into the optically thick wind regime because the continuum contribution to the radiative-driving mechanism becomes more relevant as we approach the Eddington limit, reason why the $\Gamma_\text{e,trans}$ is almost constant for different values of $\alpha$ CAK parameter.
	Hence, this transition differs from the transition based on the single-line scattering limit established by \citet{sabhahit22,sabhahit23}, where the kink in the $\dot M-\Gamma_\text{e}$ is due to the dominance of multi-scattering for the line-driven mechanism.
	This means, both transitions are connected to a different physical mechanisms at different wind conditions.
	The existence of the transition at $\Gamma_\text{e,trans}=0.5$ where continuum-driven wind becomes predominant does not contradict the existence of a limit where the line-driving needs to adopt multi-scattering.
	Indeed, the trend shown in Fig.~\ref{eta_vs_Gammae} suggests that the condition $\eta\simeq0.6$ should be satisfied at some point $\Gamma_\text{e}\gg0.5$, even though it is not possible to assure it properly because of the lack of studies adopting new $\dot M$ recipes for $\Gamma_\text{e}>0.4$ \citep{kk17,bjorklund21,alex23a}, whereas the transition Eddington factor determined by \citet{sabhahit22} is around $\sim0.4$ (see their Table~3).
	This implies, that either new winds predict a transition wind efficiency lower than the $\eta_\text{switch}$ found by keeping V01 for thin winds (thus meaning that the $\eta$ transition would depend on the mass-loss recipe adopted for low $\Gamma_\text{e}$), or the transition to thick winds might be better represented with the predominance of electron scattering for the radiative acceleration when Eddington factor goes around $\sim0.4-0.5$.
	As discussed in Sec.~\ref{selfconsistent_gamma05}, this is exactly the range of $\Gamma_\text{e}$ where self-consistent mass-loss rate shows more dispersion.
	
	Another aspect that should be explored, is the metallicity dependence.
	\citet{sabhahit23} finds that the $\eta_\text{switch}$ value where the wind becomes optically thick varies with the metallicity, given the direct dependence on $\dot M$ and terminal velocity for the wind efficiency number \citep{vink12}, and the adopted scaling for the terminal velocities a different $Z$.
	From our side, the transition based on $\Gamma_\text{e,trans}=0.5$ suggests to be invariant for different metallicities, because of its invariance for different values of the line-force $\alpha$ (which it is metallicity dependent).
	However, as seen in Sec.~\ref{selfconsistent_gamma05}, a deeper analysis of the m-CAK around $\Gamma_\text{e}>0.4$ will be required at various values of $Z$, both to evaluate the possibilities of some metallicity dependence for $\Gamma_\text{e,trans}$ but also to get $\dot M_\text{thick}(Z)$.
	Up to now, we just have the calibration done by \citet{brands22} plus the $\dot M-Z$ dependence from \citet{alex23a}, fitted only for the mass range of $25-120\,M_\odot$ and thus extrapolated up to $M_\text{zams}=200\,M_\odot$ in this work.
	
	For a more general prospective, it is important to mention the increase of the Eddington factor as the star evolves through the main sequence, with weaker winds boosting this tendency for the cases with $M_\text{zams}\gtrsim70\,M_\odot$ \citep[Fig.~A7]{alex23a}.
	The more massive the star, the faster it reaches $\Gamma_\text{e,trans}=0.5$ and therefore the shortest its fraction of MS lifetime adopting $\dot M_\text{sc}$.
	In the case of the evolution models introduced in this work, a $60\,M_\odot$ star will spend $\sim92\%$ of its MS lifetime exhibiting optically thick winds ($\sim92\%$ for the \texttt{MESA} model), in contrast with the $\sim7\%$ of the MS lifetime for a $200\,M_\odot$ star ($\sim4\%$ for the \texttt{MESA} model).
	These percentages of the lifetime are given by the location and track across the HRD and not by their $\dot M$ adopted during the optically thin regime.
	Therefore, the chosen stellar masses for the analysis in this work represent the approximated limits where optically thin and thick winds can coexist during the main sequence stage.
	For stars with $M_\text{zams}<60\,M_\odot$ the limit $\Gamma_\text{e}=0.5$ may not be reached during the MS and therefore the stars will just become classical WR after the depletion of hydrogen, as in traditional evolution models.
	On the other limit, for stars with $M_\text{zams}\ge200\,M_\odot$ it is satisfied that $\Gamma_\text{e}=0.5$ from the ZAMS and therefore the wind is fully optically thick, as also found by \citet{sabhahit23}.
	In the intermediate mass range, stars will spend one fraction of their lifetime with thin OB-type winds and reduced mass-loss as reported by \citet{grafener21}, and the other fraction with the enhanced $\dot M$ if WNh stars.
	Summarising, the new weaker winds from \citet{alex22b,alex23a} can be applied for all initial masses for the earliest stages, even for VMS, for the segment of time when $\Gamma_\text{e}<0.5$ is satisfied.

% Spectroscopic analysis
	\begin{figure*}[t!]
		\centering
		\includegraphics[width=0.41\linewidth,align=c]{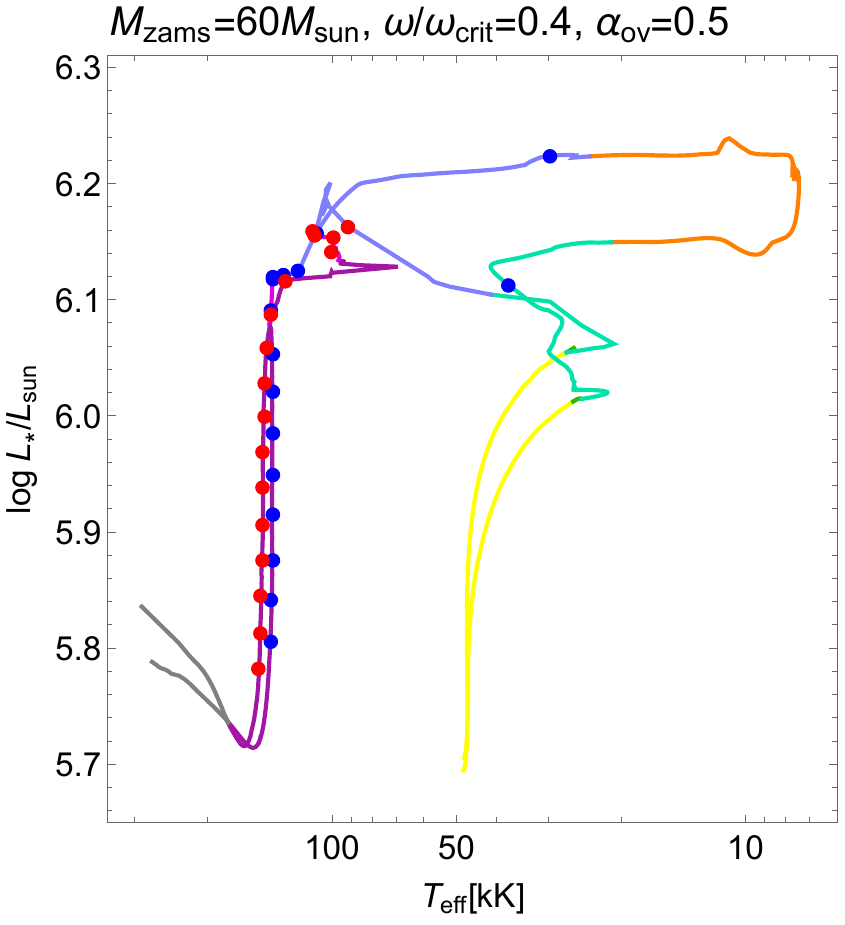}
		\hspace{2mm}
		\includegraphics[width=0.13\linewidth,align=c]{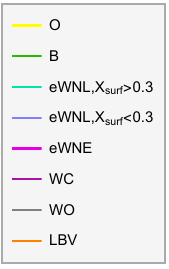}
		\hspace{2mm}		
		\includegraphics[width=0.41\linewidth,align=c]{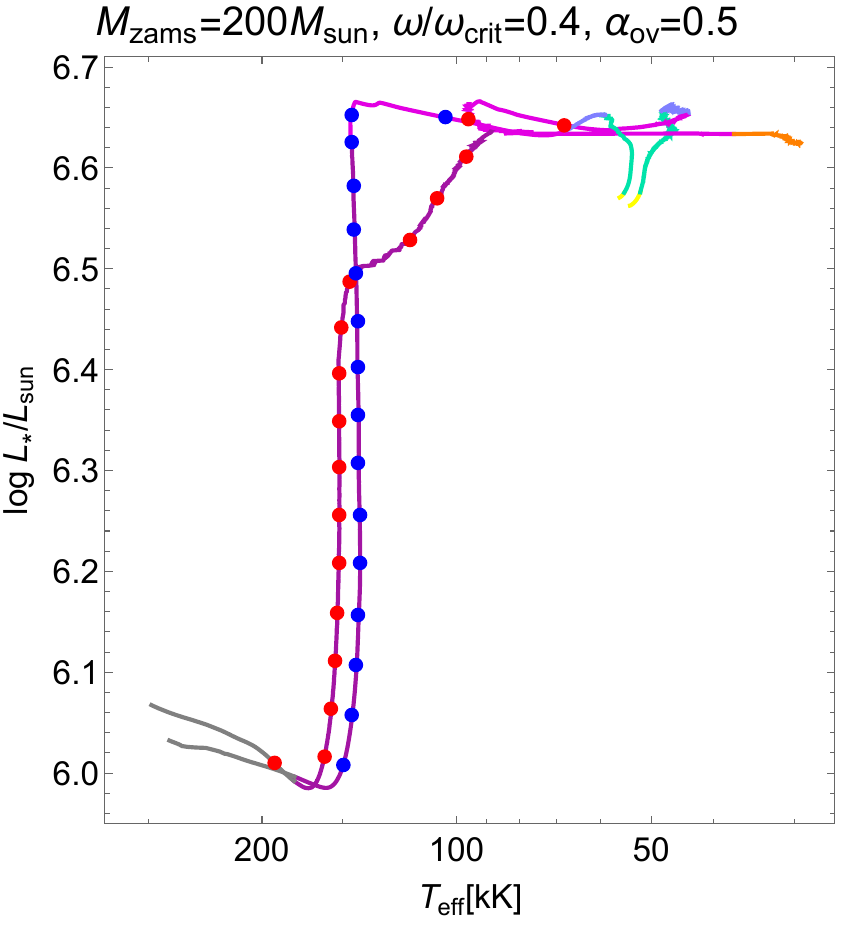}
		\caption{\small{\text{Distribution across the HRD of the evolutionary spectral types introduced in Sec.~\ref{spectroscopicanalysis}.
		The red and blue dots represents time intervals of 20\,000 years starting from the end of the H-core burning, for our \textsc{Genec} and \texttt{MESA} models respectively.}}}
		\label{spectralHRD}
	\end{figure*} 
\subsection{Spectroscopic analysis}\label{spectroscopicanalysis}
	The new mass-loss prescription, plus the new transition from thin to thick winds, modify the evolutionary tracks across the HR diagram and thus the spectroscopic phases.
	According to the output of our evolution models, we can classify the spectroscopic phases of our evolution models according to
 
	\begin{itemize}
		\item O-type: optically thin wind and $T_\text{eff}\ge26.3$ kK.
		\item B-type: optically thin wind and $T_\text{eff}<26.3$ kK.
		\item \text{eWNL}: WR star with $X_\text{surf}\ge0.05$.
		\item \text{eWNE}: WR star and $X_\text{surf}<0.05$ and $(^{12}\text{C}+^{16}\text{O})/^{4}\text{He}\le0.03$.
		\item WC: WR star wind and $(^{12}\text{C}+^{16}\text{O})/^{4}\text{He}>0.03$.
		\item WO: WR star with $\log T_\text{eff}\ge5.25$.
	\end{itemize}

	These WR boundaries are derived from the conditions settled by \citet{yusof13}, \citet{aadland22}, and \citet{martinet23}; whereas the distinction between O and B-type is taken from \citet{groh14}.
	It is important to note, that this classification is not based on spectral analysis but on the output of our theoretical models, and thus discrepancies may be observed.
	For example, we separate the evolutionary eWNL and eWNE \citet[where `e' stands from evolutionary, following][]{foellmi03} according to the hydrogen abundance as done by \citet{meynet03}, but the separation between WNL and WNE spectral subtypes is based on the ionisation levels of the surface nitrogen \citep{smith96}.
	This could suggest that all WNE stars are H-poor, albeit there are evidence of few H-rich WNE stars in the Milky Way \citep{hamann19}.
	These H-rich WNE stars however, have luminosities below $\log L_*/L_\odot\simeq5.5$ which suggest their progenitors were stars with $M_\text{zams}\simeq40\,M_\odot$ \citep{georgy12}, thus being out of the range of our analysis.
	The results are shown in Table~\ref{table_spectroscopicphases}, whereas the location of these spectral phases across the HRD is shown in Fig.~\ref{spectralHRD}.

	Even though is not a formal spectroscopic phase, we classify our star as a luminous blue variable (LBV) if the HD-limit ($L>6\times10^{5}$ and $10^{-5}L^{1/2}R>1$) is exceeded \citep{humphreys94,hurley00}, although recent works have found a lower luminosity limit of $L>3\times10^{5}$ \citep{davies20}.
	This is particularly evident for our $60\,M_\odot$ {\tt MESA} model, where the post MS expansion exceeds the HD limit and therefore the star spends $12$ kyr ($0.26\%$ of its total lifetime) as a LBV.
	Despite this short period, the star had a mass of $M_\text{pre,lbv}\simeq44.1$ $M_\odot$ and $X_\text{H}\simeq0.4$, to finish with $M_\text{post,lbv}\simeq41.9$ $M_\odot$ and $X_\text{H}\simeq0.24$, i.e., it experienced an average mass loss of $\dot M_\text{lbv}\simeq1.8\times10^{-4}$ $M_\odot$~yr$^{-1}$ ($\log\dot M_\text{lbv}\simeq-3.7$).
	Such value is close to the adopted $\dot M_\text{lbv}$ by \citet{belczynski10a}, but still it comes from the application of the wind recipes of \citet{vink01} and \citet{dejager88} in a region of the HRD beyond the HD limit.
	Models for eruptive mass loss \citep{quataert16,cheng24}, more in agreement with the nature of LBVs, would be more require to better represent the evolution during this phase.
	In parallel, our $60\,M_\odot$ \textsc{Genec} does not become a LBV because its more chemically homogeneous structure avoids any radial expansion after the H-depletion.
	Therefore, despite that the final masses tabulated for both codes are quite similar, the prediction about the existence or not of the LBV phenomenon strongly depends on the calculated mixing during the main sequence stage.
	In opposition, our $200\,M_\odot$ \textsc{Genec} evolution track moves redwards, crossing the HD limit, between its phases WNE and WC.
	This reinforces the uncertain relation of the LBV phenomenon with the evolution of massive stars because, if we consider that redwards expansion as a physical phenomenon instead of a numerical artefact, it would mean that LBV events \text{could} occur even after the star becomes a hydrogen depleted WN star.
	However, this has never been observationally found, and then the idea of a numerical artefact due to a drastic change in the opacity due to the removal of the last traces of hydrogen looks more compelling.

\begin{figure}[t!]
		\centering
		\includegraphics[width=\linewidth]{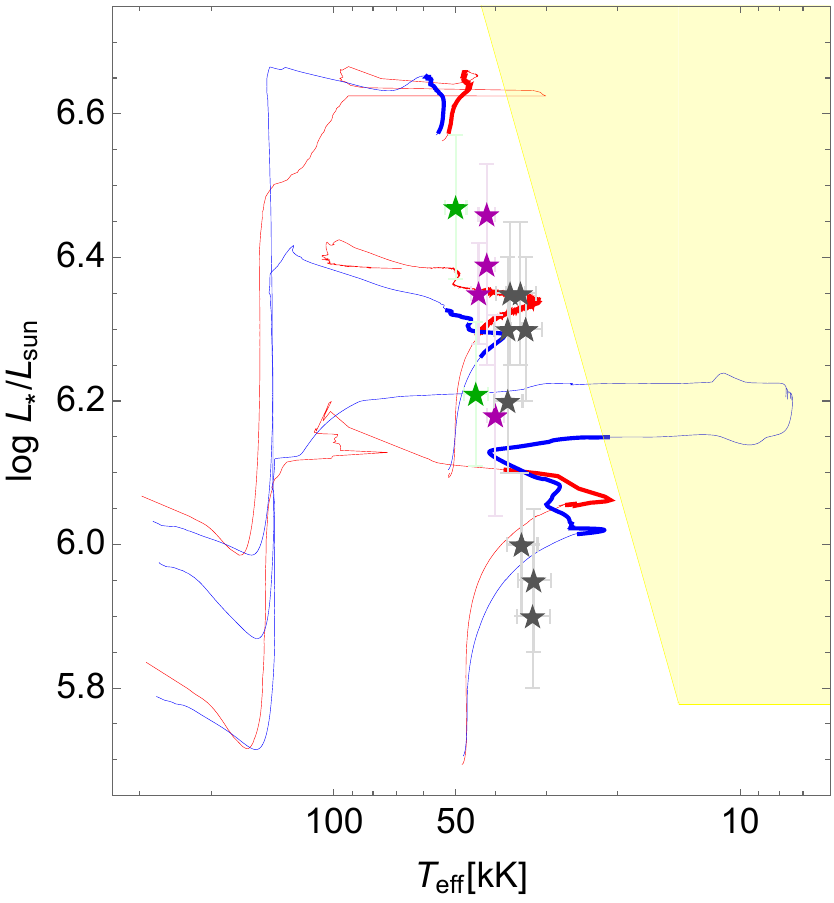}
		\caption{\small{Evolutionary \textsc{Genec} and {\tt MESA} tracks for $60$ and $200\,M_\odot$ from Fig.\ref{fullHRD} and additional $100\,M_\odot$ models, where the thick solid lines represent the segments where the wind is optically thick ($\Gamma_\text{e}\ge0.5$) and $X_\text{surf}\ge0.3$.
		The gray stars correspond to the sample of WNh stars from \citet{martins08}, the purple stars to the sample of \citet{crowther10} and the green stars to the sample of \citet{hamann19}, tabulated in Table~\ref{WNh_martinsandhamann}.
		Yellow shadowed area correspond to the HRD region where the HD limit is exceeded.}}
		\label{HRD_Martins}
	\end{figure}

	Besides the standard classification for WR stars, in Table~\ref{table_spectroscopicphases} \text{and Fig.~\ref{spectralHRD}} we also separate the WNL stars with $X_\text{H}>0.3$.
	Given that typical stellar evolution models enter into the WR phase when either $X_\text{H}<0.3$ \citep{ekstrom12} or $X_\text{H}<0.4$ \citet{brott11}, this distinction is made to incorporate the WNh stars whose optically thick winds are because of the proximity to the Eddington limit before the H-depletion, according to the criterion $\Gamma_\text{e}\ge0.5$ established in Sec.~\ref{eddingtonlimit}.
	Thus we discern from the WNL stars which are result of the `classical' evolution, where the star becomes a WR due to the drift bluewards after the LBV episodes during the He-core burning stage \citep{conti75,groh14}.
	Indeed, WNh stars with $X_\text{H}\ge0.3$ are constrained to a very narrow band of temperature in the HRD, $\log T_\text{eff}\simeq4.5-4.7$ \citep{martins08,crowther10,hamann19}, whereas WNL stars with $X_\text{H}\le0.3$ are spread across a wider range of temperatures \citep{martins23}.
	To illustrate better this point, we select the Galactic WNh stars (with $X_\text{H}\ge0.3$) from the Arches cluster \citep{martins08}, from the NGC 3603 cluster \citep{crowther10}, and from the local neighbourhood \citep{hamann19}, and we tabulated them in Table~\ref{WNh_martinsandhamann}.
	In addition, we plot our \textsc{Genec} and {\tt MESA} models for $60$ and $200$ $M_\odot$, highlighting the segments where $\Gamma_\text{e}\ge0.5$ and $X_\text{H}\ge0.3$.
	The results are shown in Fig.~\ref{HRD_Martins}.
	Despite the limited differences between the evolution tracks, our prescription adequately describes the narrow band where WNh exist, thus meaning that WNh are the product of stars born with $M_\text{zams}\gtrsim60\,M_\odot$ which produce optically thick winds before the H-depletion due to their proximity to the Eddington limit.
	In the middle we append models for $100\,M_\odot$, showing that the thick band also fits the WNh of our sample.

	For the $200\,M_\odot$ cases the observational diagnostics are more uncertain.
	For instance, the stars tabulated in Table~\ref{WNh_martinsandhamann} are also the brightest WNh stars in the Milky Way, and none of them reaches the luminosity of $\log L_*/\log L_\odot\gtrsim6.55$.
	The early approaching to the Eddington limit from the very beginning of the main sequence implies that the O-type phase in our models is very short, and therefore the star will spend the largest percentage of their lifetime (almost a $\sim90\%$ for both models) as a WNh.
	This is a remarkable difference with respect to previous studies, where very massive stars still spent the majority of their lifetime as O-type stars until they remove more than the half of their initial surface H abundance \citep{martinet23}.
	In other words, our models predict that extremely luminous Galactic stars have more probabilities of being WNh than O-type.
	The star W49-\#2, with $\log L_*/L_\odot\simeq6.64\pm0.25$ (one of the brightest stars known in the Milky Way) is classified as O2-3.5If* \citep{wu16} but quite close of being a Of/WN star based on the spectral classification of \citet{crowther11}.
	In any case, the high stellar mass ($\sim250\,M_\odot$) estimated by \citet{wu16} implies $\Gamma_\text{e}\simeq0.35-0.46$ (depending on the $X_\text{H}$ to be considered in Eq.~\ref{logGammaEdd}), thus not being close to the Eddington limit enough to develop optically thick winds.

	Beyond WNh stars, our new evolution models for $200\,M_\odot$ also predicts the existence of H-poor \text{WR} stars with luminosities larger that $\log L_*/L_\odot>6.0$, same as the rotating models of $180\,M_\odot$ and $250\,M_\odot$ from \citet{martinet23}.
	If we sum up the lifetime of eWNE, WC and WO spectroscopic phases from Table~\ref{table_spectroscopicphases}, we find that the our $200\,M_\odot$ models will spent a $\sim14\%$ of their lifetime according to the \textsc{Genec} model ($\sim13\%$ in the case of \texttt{MESA}), very close to the $\sim11\%$ for the same phases predicted by them.
	Such small discrepancy can be easily explained by our choice of setting eWNE for $X_\text{surf}\le0.5$ \citep[from]{yusof13} instead of $\le10^{-5}$, which implies that eWNE will form around the $\sim3.5\%$ of the total lifetime (either H-rich or H-poor) in contrast with the absence of eWNE phase for $200\,M_\odot$ from \citet{martinet23}.
	Nonetheless, only two WN (WR 18 and WR 37) stars with $X_\text{H}=0$ satisfies $\log L_*/L_\odot>6.0$ in the Galactic catalog of \citet{hamann19}; whereas for WC type, WR 126 is the only WN/WC star with $\log L_*/L_\odot>6.0$ in the Milky Way \citep{sander19}.
	However, the trace of the red and blue dots from Fig.~\ref{spectralHRD} confirms that eWNE is a very short period, and these stars quickly drops their luminosities from $\log L_*/L_\odot\ge6.6$ to $\simeq6.0$ in just $\sim200$ kyr.
	Even though the mentioned catalogs still cover a small region of the Milky Way, the absence of WNE and WC stars with $\log L_*/L_\odot\ge6.1$ (which should be evolved from models born with masses $\simeq60$) should suggest that Eq.~\ref{mdotthick} might be underestimating the actual value of the mass-loss rate for the most massive H-rich WN stars, as a consequence of the extrapolation up to $200\,M_\odot$ mentioned in Sec.~\ref{comparisonwitheta}.
	Even though it is still possible that we might have been underestimating the actual value of the mass-loss rate for the most massive H-rich WN stars as a consequence of the extrapolation up to $200\,M_\odot$ mentioned in Sec.~\ref{comparisonwitheta} (and thus overpredicting the existence of H-poor WR stars at larger luminosities), it is important to mention that our Galactic WR catalogs are still quite constrained to our solar neighbourhood.
	The study and discoveries of new sources at the edge of the known stellar luminosities, together with less uncertain data, is beyond the scope of this paper.

	\begin{table}[t!]
		\centering
		\caption{WNh stars included in Fig.~\ref{HRD_Martins}.}
%		\resizebox{\linewidth}{!}{
		\begin{tabular}{cc|cccccccccccccc}
			\hline
			\hline
			Star & Sp. Type & $T_*$ & $L_*$ & $X_\text{H}$ & Ref.\\
			& & [kK] & [$L_\odot$] & mass frac. & \\
			\hline
			B1 & WN8-9h & 32.2 & 5.95 & 0.70 & 1\\
			F1 & WN8-9h & 33.7 & 6.30 & 0.70 & 1\\
			F2 & WN8-9h & 34.5 & 6.0 & 0.41 & 1\\
			F4 & WN8-9h & 37.3 & 6.3 & 0.38 & 1\\
			F7 & WN8-9h & 33.7 & 6.30 & 0.45 & 1\\
			F9 & WN8-9h & 36.8 & 6.35 & 0.70 & 1\\
			F12 & WN8-9h & 37.3 & 6.20 & 0.55 & 1\\
			F14 & WN8-9h & 34.5 & 6.00 & 0.70 & 1\\
			F16 & WN8-9h & 32.4 & 5.90 & 0.70 & 1\\
			A1a & WN6h & 42.0 & 6.39 & 0.60 & 2\\
			A1b & WN6h & 40.0 & 6.18 & 0.70 & 2\\
			B & WN6h & 42.0 & 6.46 & 0.60 & 2\\
			C & WN6h & 44.0 & 6.35 & 0.70 & 2\\
			WR 24 &  WN6ha-w & 50.1 & 6.47 & 0.44 & 3\\
			WR 87 & WN7h & 44.7 & 6.21 & 0.40 & 3\\
			\hline
		\end{tabular}
		\tablebib{(1)~\citet{martins08}; (2)~\citet{crowther10}; (3)~\citet{hamann19}.}
		\label{WNh_martinsandhamann}
	\end{table}

%_____CONCLUSIONS_______________________________________________________________________________
\section{Summary and conclusions}\label{conclusions}
	In Paper I \citep{romagnolo24} we explored the expected final masses prior to the core collapse, when we calculate stellar evolution updating the mass-loss prescription.
	In this work, we confirm the importance of these new wind prescriptions, by analysing its impact over the expected spectroscopic stages (OB-type, WNh and classical WR).

	Regardless of the unification of the overshooting ($\alpha_\text{ov}=0.5)$, the rotation ($\omega/\omega_\text{crit}=0.4$), the angular momentum transport (Tayler-Spruit dynamo), the convective boundary criterion (Ledoux), and the line-driven mass-loss prescriptions for both codes \textsc{Genec} and \texttt{MESA}, there are important differences between the calculated evolutionary tracks.
	For instance, \textsc{Genec} computes a more prominent rotational mixing during the main sequence phase, leading to a more homogeneous stellar structure at the H-core depletion in comparison with \texttt{MESA}.
	This is particularly evident for our $60\,M_\odot$ models, where our less homogeneous model largely expands and crosses the HD limit before the He-ignition.
	Such discrepancy between both models can be interpreted as a boundary in the mass range below which stars normally expands, and over which stars keep a moderate size because of the strong outflows \citep{romagnolo23}.
	For the $200\,M_\odot$ models, due to the large size of the convective core the differences between both evolution codes are less relevant, and only the treatment for superadiabacity during the H-core burning phase is a matter of debate.
	
	Concerning stellar masses, both \textsc{Genec} and \texttt{MESA} codes show similar results at the end of each core burning process, with the most remarkable mass losses during the post-MS stages.
	Models with $M_\text{zams}=60\,M_\odot$ end the H-core burning with final masses of $42.1$ and $44.5\,M_\odot$ respectively, subsequently experiencing strong outflows during He-core burning stages to finish with $16.1\,M_\odot$ at the C-core depletion.
	This is a striking discrepancy with the $M_\text{bh}\simeq30\,M_\odot$ found at $M_\text{zams}=60\,M_\odot$ by \citet{bavera23}, where the rotational effects were not considered, or with \citet{vink24}, where the WR stage is absent for $M_\text{zams}=40\,M_\odot$.
	For our $200\,M_\odot$ case, where the main sequence is almost fully dominated by thick winds, both \textsc{Genec} and \texttt{MESA} models predict a final mass of $\sim100\,M_\odot$ at the end of the H-core burning stage, finishing later with $M_\text{fin}\simeq24.6\,M_\odot$ at the C-core depletion.
	This result is also largely discrepant with other studies such as \citet{yusof13} or \citet{sabhahit22,sabhahit23}, who found that a star with $M_\text{zams}=200\,M_\odot$ would loss almost the $\sim80\%$ of its mass during the H-burning stage only.
	Despite the discussion introduced in Section~\ref{selfconsistent_gamma05}, such difference in the calculated mass remnants is not produced due to the approach of the transition from thin to thick winds close to the Eddington factor, but due to the lower recipes for $\dot M$ for both thin –\citep{alex23a} instead of V01– and thick –\citet{bestenlehner20} instead of \citet{vink11}– wind regimes.

	However, the most remarkable result is that our transition between the optically thin winds of OB-type stars and the thick winds of WNh stars based on the dominance of electron scattering over line-driving, results in the prediction of a band in the HRD where stars are expected to develop optically thick winds but with large hydrogen fractions at the surface ($X_\text{surf}\gg0.3$).
	This band corresponds to effective temperatures between $\log T_\text{eff}\simeq4.5-4.7$, in agreement with the spectroscopic diagnostics of \citet{martins08}, \citet{crowther10}, and \citet{hamann19}.
	The basis for the proposed switch of wind regime at $\Gamma_\text{e,trans}=0.5$ follows the formulation of \citet{bestenlehner20} based on CAK theory, and it considers that at early evolutionary stages the $\dot M$ for regular massive stars (i.e., not VMS) is $\sim3$ times lower than the traditionally assumed for V01.
	This statement is compatible with the idea of having WNh stars born from $M_\text{zams}\simeq60\,M_\odot$, whereas we find reduced $\dot M$ from hydrodynamic analysis on winds for the same mass range \citep{sander17,alex21}.
	As we escalate in the mass range, the fraction of MS time at which the stellar wind is optically thin becomes shorter, and therefore the wind of very massive stars posses enhanced values of mass-loss rates because they spent larger fractions of their MS dominated by thick winds.
	Stars at $M_\text{zams}\gtrsim200\,M_\odot$ are already dominated by thick winds from the very beginning of their lifetimes, similar as found by \citet{sabhahit23}.
	
	The implementation of $\Gamma_\text{e}=L_*\kappa_\text{e}/4\pi cGM_*$ instead of $\eta=\dot M\varv_\infty/(L_*/c)$ is explicitly independent of the adopted $\dot M$ recipe for thin winds, thus being less dependent on the mass-loss history before the O/WNh transition.
	The value $\log\dot M_\text{thick}(\Gamma_\text{e,trans}=0.5)\sim-4.9$ is in rule with the so-called `model-independent' mass-loss transition from \citet{vink12}, even though there is a difference of $\sim0.5$ dex for $\dot M_\text{thin}$ at larger values of $\Gamma_\text{e}$.
	This disagreement could be avoided by putting the value of $\Gamma_\text{e,trans}$ down to $\sim0.24$, but for the purposes of this study we keep $\Gamma_\text{e,trans}=0.5$ to preserve the physical meaning of an equilibrium between the line-driven and the electron scattering to the total radiative acceleration.
	To better connect the new recipes for O-type stellar winds with the enhanced winds of WNh stars it should be required to expand the self-consistent wind calculations of \citet{kk17} or \citet{alex21} for $\Gamma_\text{e}\gg0.4$, exploring the evolution of the resulting $\eta$ from ZAMS to the O/WNh transition.

	This work focuses on calculations applicable to stars with solar metallicity and in earlier stages of evolution.  
	However, recent observations reveal a significant population of Wolf-Rayet stars (WNh) in young and metal-poor clusters like R136 \citep{brands22} and 30 Dor \citet{ramirezaguledo17}, both located in the LMC.
	This necessitates extending our procedures and diagnostics to encompass very massive stars with metallicities as low as $Z\lesssim0.006$, as suggested by \citet{martins22}. 
	Additionally, these upgrades in mass-loss prescriptions need to be applicable to stars with $T_\text{eff}\le30$ kK.
	Evaluating approaches like those proposed by \citet{krticka24} for B-supergiants is a promising avenue for future work, together with incorporating more detailed prescriptions for the stellar interiors based on the state-of-the-art asteroseismology \citep{pedersen21}, besides considering that the constrain of stellar parameters and mass-loss rates of massive stars in metal poor environments is still an active field of research \citep{sander24,verhamme24,backs24,gomezgonzalez24}.

%_____AGRADECIMIENTOS___________________________________________________________________________
\begin{acknowledgements}
	We thank to the anonymous second referee for their valuable comments and feedback, together with to Joachim Bestenlehner, Sylvia Ekström, and Michel Curé for the fruitful discussions.
	Authors acknowledge support from the Polish National Science Center grant Maestro (2018/30/A/ST9/00050).
	The authors also thank the \textsc{Genec} and \texttt{MESA} communities at large for their helpful feedback on the models creation.
	Computations for this article have been performed using the computer cluster at CAMK PAN.
	ACGM thanks the support from project 10108195 MERIT (MSCA-COFUND Horizon Europe).
	We dedicate this paper to Krzysztof Belczyński, who contributed to this research before his untimely passing on 13th January 2024.
\end{acknowledgements}

%_____BIBLIOGRAFÍA_______________________________________________________________________________
\bibliography{evol60and200paper.bib} % your references Yourfile.bib
\bibliographystyle{aa} % style aa.bst

\begin{appendix}
%_____EFFECTS FROM INNER STRUCTURE
\section{Effects of TS dynamo for angular momentum transport}
	\begin{table*}[t!]
		\centering
		\caption{Table of final parameters for our \textsc{Genec} evolution models, comparing the new inner structure (Tayler-Spruit dynamo and Ledoux convective criterion) with the old inner structure (no-TS dynamo and Schwarzchild convective criterion).}
		\resizebox{\linewidth}{!}{
		\begin{tabular}{ccc|cccccc|cccccc|cccccc}
			\hline
			\hline
			\multicolumn{3}{c}{Initial parameters} & \multicolumn{6}{c}{End of H-core burning} & \multicolumn{6}{c}{End of He-core burning} & \multicolumn{6}{c}{End of C-core burning}\\
			$M_\text{zams}$ & \multicolumn{2}{l}{inn struct} & $M_\text{tot}$ & $M_\text{core}$ & $R_\text{max}$ & $\tau_\text{H}$ & $Y_\text{surf}$ & $\Gamma_\text{e}$ & $M_\text{tot}$ & $M_\text{core}$ & $R_\text{max}$ & $\tau_\text{He}$ & $Y_\text{surf}$ & $\Gamma_\text{e}$ & $M_\text{tot}$ & $M_\text{core}$ & $R_\text{max}$ & $\tau_\text{C}$ & $Y_\text{surf}$ & $\Gamma_\text{e}$\\
			$[M_\odot]$ & & & $[M_\odot]$ & $[M_\odot]$ & $[R_\odot]$ & [Myr] & & & $[M_\odot]$ & $[M_\odot]$ & $[R_\odot]$ & [Myr] & & & $[M_\odot]$ & $[M_\odot]$ & $[R_\odot]$ & [Myr] & &\\
			\hline
			60 & \multicolumn{2}{l}{TS dyn, Led} & 44.8 & 37.6 & 44.8 & 4.425 & 0.853 & 0.599 & 17.3 & 12.2 & 7.3 & 4.784 & 0.051 & 0.508 & 16.7 & 12.6 & 0.84 & 4.789 & 0.037 & 0.675\\
			60 & \multicolumn{2}{l}{mer cur, Sch} & 48.1 & 40.2 & 18.9 & 4.507 & 0.974 & 0.572 & 16.5 & 11.5 & 16.1 & 4.880 & 0.052 & 0.488 & 15.9 & 12.6 & 0.81 & 4.885 & 0.037 & 0.656\\
			\hdashline
			200 & \multicolumn{2}{l}{TS dyn, Led} & 100.2 & 87.9 & 38.2 & 2.525 & 0.972 & 0.691 & 25.5 & 18.3 & 10.4 & 2.841 & 0.034 & 0.594 & 24.4 & 19.2 & 1.1 & 2.845 & 0.023 & 0.733\\
			200 & \multicolumn{2}{l}{mer cur, Sch} & 106.2 & 94.5 & 42.1 & 2.569 & 0.976 & 0.698 & 25.6 & 18.7 & 9.9 & 2.886 & 0.034 & 0.592 & 24.6 & 19.4 & 1.1 & 2.890 & 0.023 & 0.739\\
			\hline
		\end{tabular}}
		\label{table_TSLed_vs_def}
	\end{table*}
	\begin{figure*}[t!]
		\centering
		\includegraphics[width=0.4\linewidth]{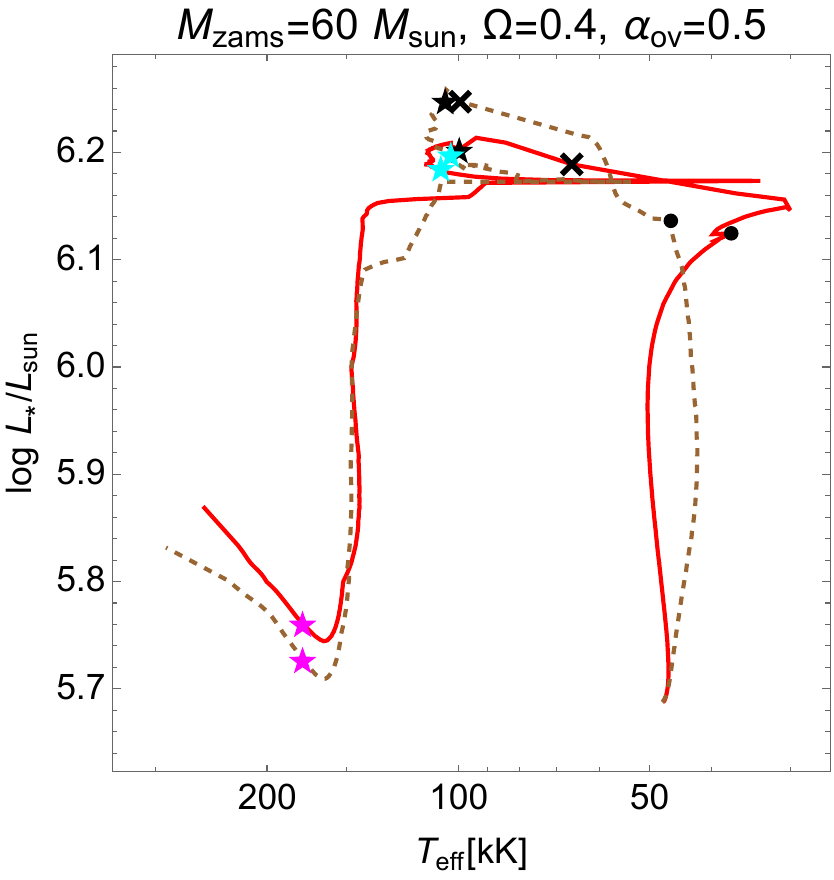}
		\hspace{1cm}
		\includegraphics[width=0.4\linewidth]{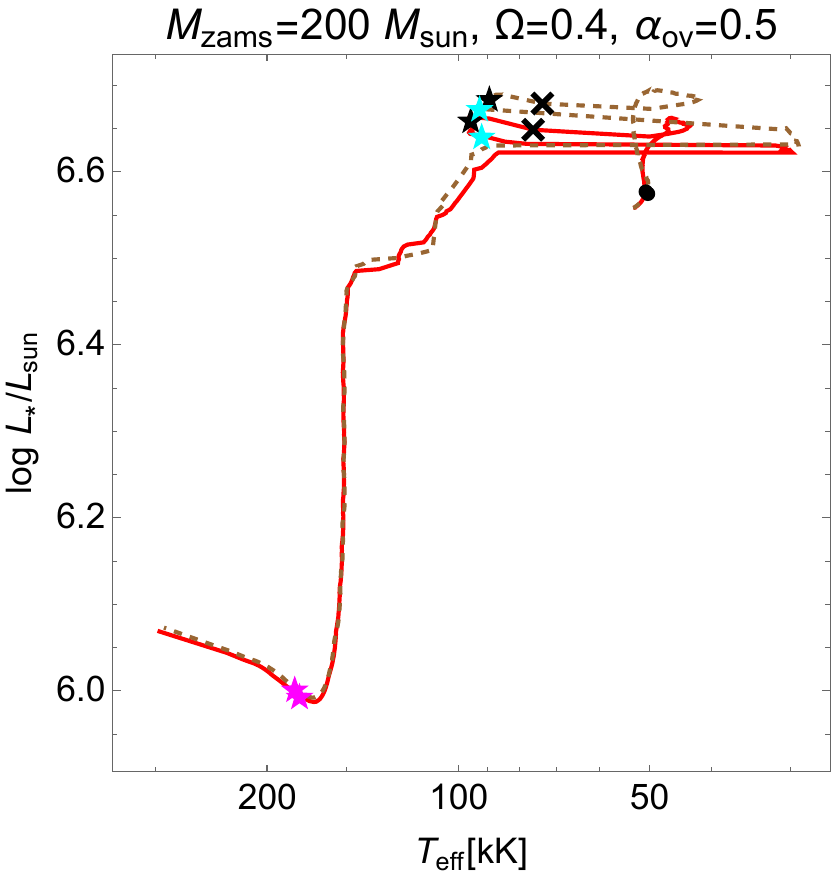}
		\caption{\small{HR diagram for our \textsc{Genec} evolution models, comparing new inner structure (Tayler-Spruit dynamo and Ledoux convective criterion, red solid lines) with old inner structure (no-TS dynamo and Schwarzchild convective criterion, dashed brown lines).}}
		\label{HRD_TSLed_vs_def}
	\end{figure*}
	\begin{figure*}[t!]
		\centering
		\includegraphics[width=0.37\linewidth]{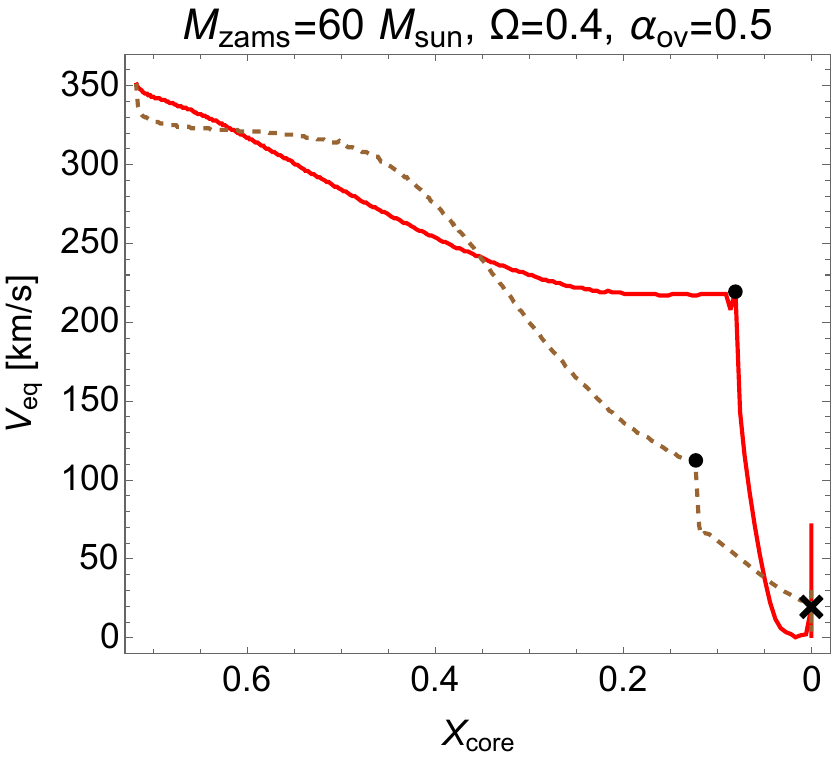}
		\hspace{1.5cm}
		\includegraphics[width=0.37\linewidth]{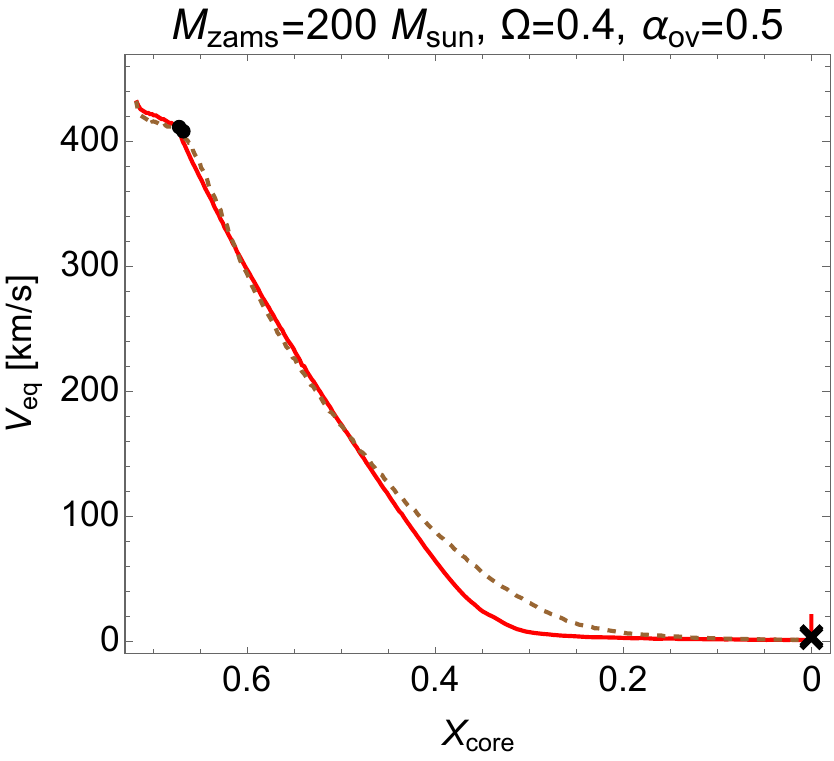}
		\caption{\small{Rotational velocity during the MS for our \textsc{Genec} evolution models, comparing new inner structure (Tayler-Spruit dynamo and Ledoux convective criterion, red solid lines) with old inner structure (no-TS dynamo and Schwarzchild convective criterion, dashed brown lines).}}
		\label{vrot_TSLed_vs_def}
	\end{figure*}
	\begin{figure*}[t!]
		\centering
		\includegraphics[width=0.37\linewidth]{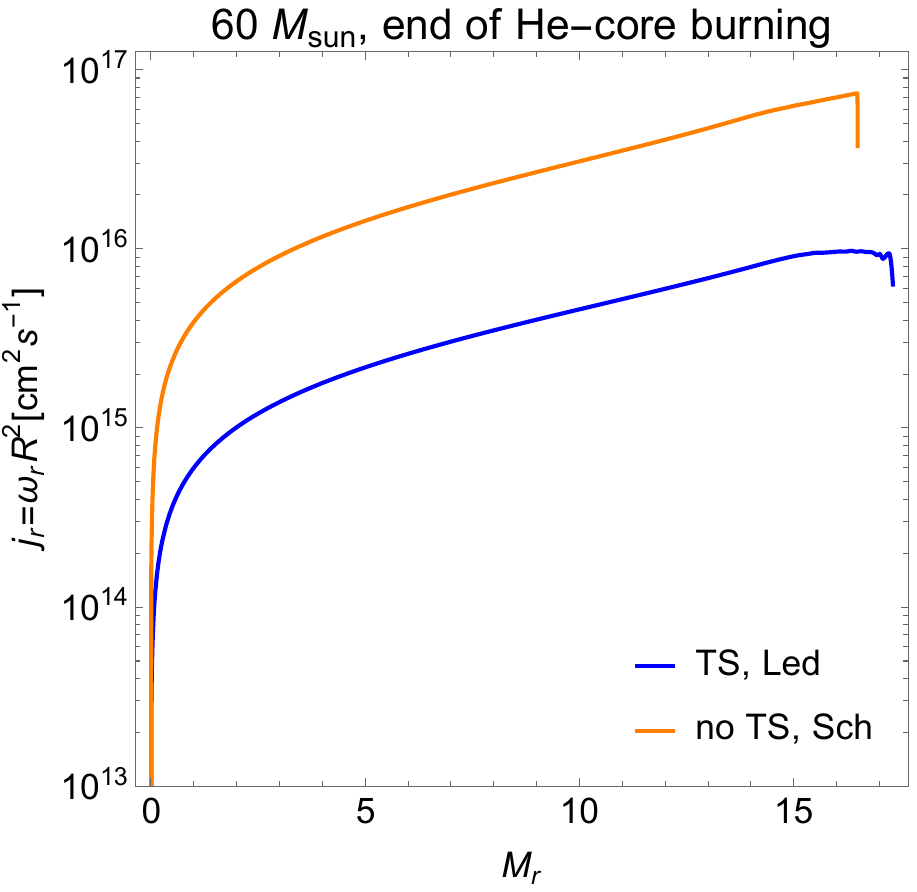}
		\hspace{1.5cm}
		\includegraphics[width=0.37\linewidth]{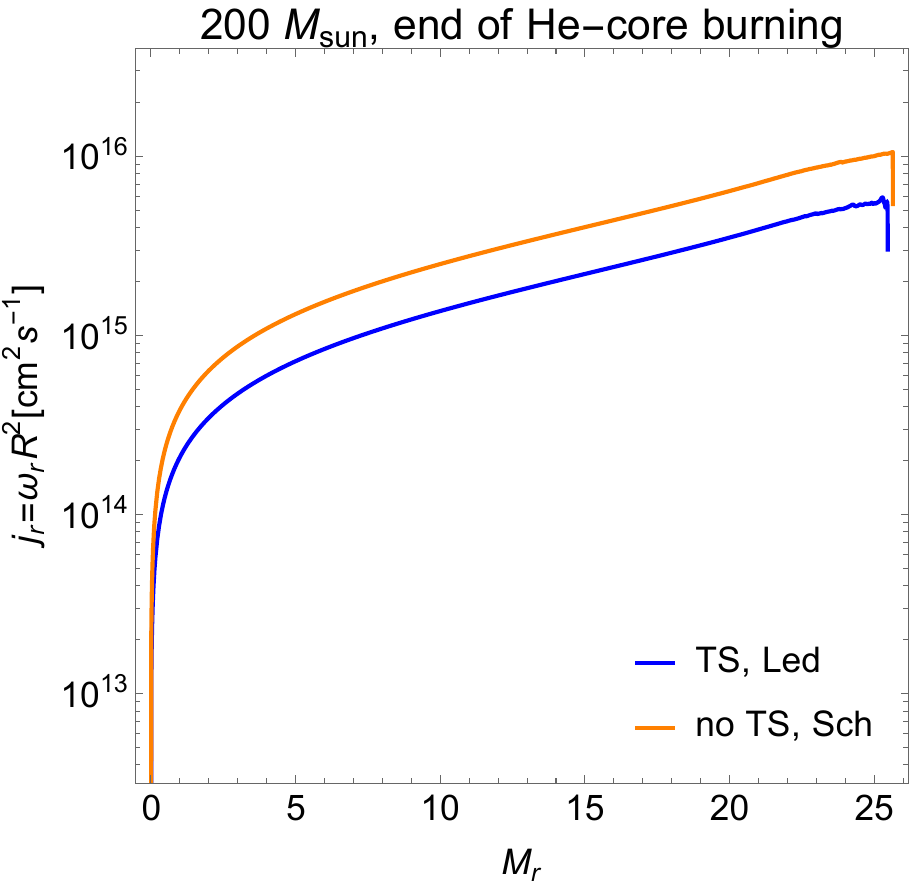}
		\caption{\small{Specific angular momentum at the end of the He-core burning for our \textsc{Genec} evolution models, comparing new inner structure (Tayler-Spruit dynamo and Ledoux convective criterion, blue) with old inner structure (no-TS dynamo and Schwarzchild convective criterion, orange).}}
		\label{angmom_TSLed_vs_def}
	\end{figure*}
	
	For our \textsc{Genec} models, we include Tayler-Spruit dynamo \citep{tayler73,spruit02,heger05} for the treatment of the angular momentum transport, besides adopting Ledoux instead of Schwarzschild criterion for the treatment of convective layers \citep{martinet23}.
	The evolution during the MS phase is basically not affected by differences in convection treatment, and just for the H-core burning the Ledoux criterion improves the agreement between models and observations for the surface abundances of BSGs evolving back from RSGs \citep{georgy14}.
	For stars with $M_\text{zams}\ge60\,M_\odot$ (as in this study), the impact of the different convective criterion is almost negligible because mass loss become the main driver of the evolution \citep{sibony23}.
	
	In Fig.~\ref{HRD_TSLed_vs_def} we show the evolution tracks of our models shown in Section~\ref{mainresults}, compared to models with the same input but adopting the original inner structure of \textsc{Genec}.
	The initial rotation this time follows the framework $\Omega=\varv/\varv_\text{crit}=0.4$ (equivalent to $\omega/\omega_\text{crit}\simeq0.58$, see Eq.~\ref{vvcrit}).
	The evolution of the rotational speed during the main sequence is shown in Fig.~\ref{vrot_TSLed_vs_def}, whereas the final parameters between both cases are displayed in Table~\ref{table_TSLed_vs_def}.

	As expected, the differences between adopting the new or the old inner structure are minimal, and are reduced to braking in the stellar rotation since the ZAMS to the TAMS.
	TS dynamo produces a more efficient transport of angular momentum from the core to the surface, and therefore the stellar surface will rotate faster than in the case without TS.
	This impact is stronger for the $60\,M_\odot$ and almost negligible for the $200\,M_\odot$ models, because of the larger convective core of VMS.
	
	Regardless of the HRD tracks, TS dynamo actually affects the stellar inner structure.
	Fig.~\ref{angmom_TSLed_vs_def} shows how the specific angular momentum $j_r=\omega_iR^2$ at the end of the He-core burning stage is more depleted when TS dynamo is incorporated, due to the more efficient angular momentum transport.
	Again, the influence is more prominent for the case with the smaller convective core.
	Even though, adoption of TS dynamo have an impact on the expected spin of the resulting black hole \citep{belczynski20b}, for the analysis of the evolution of stars with $M_\text{zams}\ge60\,M_\odot$ its impact is not relevant enough.
\end{appendix}

\end{document}